\documentclass[acmtog]{acmart}

\acmSubmissionID{1260}

\setcopyright{acmlicensed}
\acmJournal{TOG}
\acmYear{2025} 
\acmVolume{44} 
\acmNumber{6} 
\acmArticle{273} 
\acmMonth{12}
\acmDOI{10.1145/3763292}

\usepackage{amsmath, amsthm, amsfonts}
\usepackage{graphicx}
\usepackage{multirow}
\usepackage{color}
\usepackage{enumitem}

\definecolor{green}{rgb}{0, 0.5, 0}
\definecolor{orange}{rgb}{0.6, 0.3, 0.1}
\definecolor{red}{rgb}{1.0, 0.0, 0.0}
\definecolor{teal}{rgb}{0.0, 0.4, 0.4}
\definecolor{purple}{rgb}{0.65,0,0.65}
\definecolor{saffron}{rgb}{0.8,0.55,0.1}
\definecolor{turquoise}{rgb}{0.0,0.5,0.5}
\definecolor{brown}{rgb}{0.5, 0.16, 0.16}
\definecolor{brickred}{rgb}{.6, .2 .1}
\definecolor{coral}{rgb}{1,0.45,0.33}

\newcommand{\eg}{{\textit{e.g., }}}
\newcommand{\ie}{{\textit{i.e., }}}

\newcommand{\ignore}[1]{}

\usepackage{subfigure}
\usepackage{wrapfig}
\usepackage[autolanguage]{numprint}

\begin{document}
	
\title{Aerial Path Planning for Urban Geometry and Texture Co-Capture}	

\author{Weidan Xiong}
\email{xiongweidan@gmail.com}
\affiliation{
	\institution{CSSE, Shenzhen University}
	\country{China}	
}

\author{Bochuan Zeng}
\email{zbc8301@gmail.com}
\affiliation{
	\institution{CSSE, Shenzhen University}
	\country{China}	
}

\author{Ziyu Hu}
\email{huziyubm05@gmail.com}
\affiliation{
	\institution{CSSE, Shenzhen University}
	\country{China}	
}

\author{Jianwei Guo}
\email{gjianwei.000@gmail.com}
\affiliation{
	\institution{Beijing Normal University}
	\country{China}	
}

\author{Ke Xie}
\email{ke.xie.siat@gmail.com}
\affiliation{
	\institution{CSSE, Shenzhen University}
	\country{China}	
}

\author{Hui Huang}
\email{hhzhiyan@gmail.com}
\authornote{Corresponding author: Hui Huang (hhzhiyan@gmail.com)}
\affiliation{
	\institution{CSSE, Shenzhen University}
	\country{China}
}

\renewcommand\shortauthors{W. Xiong, B. Zeng, Z. Hu, J. Guo, K. Xie, and H. Huang}

\begin{abstract}
Recent advances in image acquisition and scene reconstruction have enabled the generation of high-quality structural urban scene geometry, given sufficient site information. However, current capture techniques often overlook the crucial importance of texture quality, resulting in noticeable visual artifacts in the textured models. 
In this work, we introduce the urban \textit{geometry and texture co-capture} problem under limited prior knowledge before a site visit.
The only inputs are a 2D building contour map of the target area and a safe flying altitude above the buildings. 
We propose an innovative aerial path planning framework designed to co-capture images for reconstructing both structured geometry and high-fidelity textures. To evaluate and guide view planning, we introduce a comprehensive texture quality assessment system, including two novel metrics tailored for building facades. 
Firstly, our method generates high-quality vertical dipping views and horizontal planar views to effectively capture both geometric and textural details. A multi-objective optimization strategy is then proposed to jointly maximize texture fidelity, improve geometric accuracy, and minimize the cost associated with aerial views. 
Furthermore, we present a sequential path planning algorithm that accounts for texture consistency during image capture. 
Extensive experiments on large-scale synthetic and real-world urban datasets demonstrate that our approach effectively produces image sets suitable for concurrent geometric and texture reconstruction, enabling the creation of realistic, textured scene proxies at low operational cost. 
\end{abstract}


\begin{teaserfigure}
  \centering
  \includegraphics[width=\linewidth]{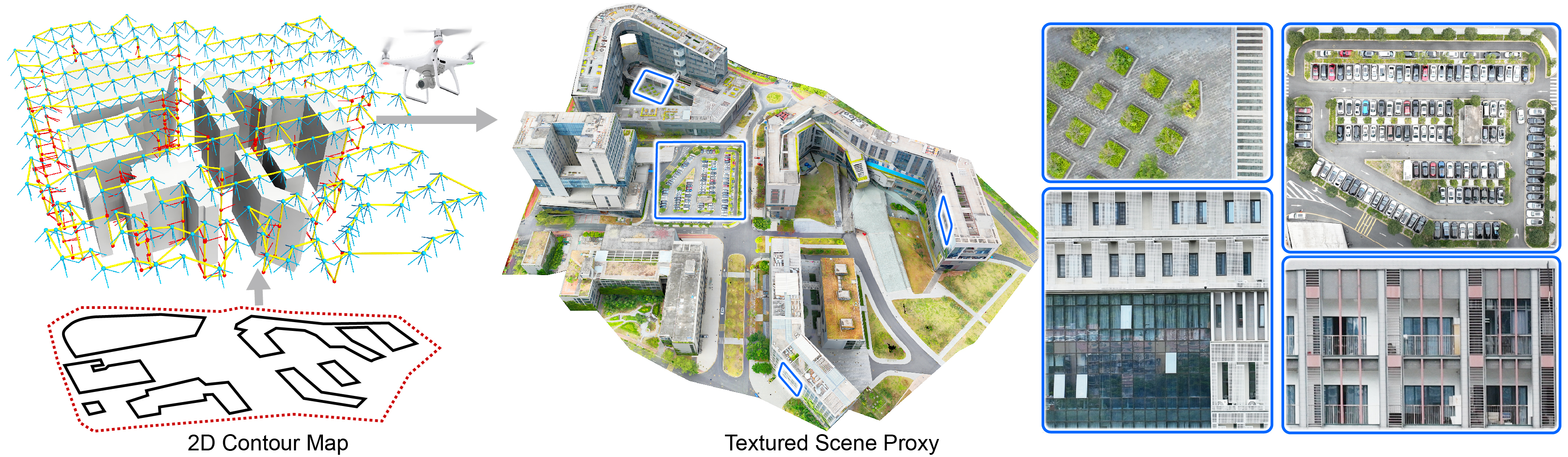}
  \caption{Given a 2D map of a large-scale complex region, our framework can plan an aerial path offsite guiding a single UAV that co-captures a collection of photos (\numprint{11710} RGB images) for reconstructing both structural geometry and high-quality textures.}
  \label{fig:teaser}
\end{teaserfigure}

\maketitle

\section{Introduction}
\label{sec:intro}

The digitization of real-world urban environments has been studied over the years~\cite{musialski2013survey}. Aerial photogrammetric sensing using unmanned aerial vehicles (UAVs) has emerged as an effective means for capturing dense geometry of large-scale urban scenes~\cite{smith2018aerial,koch2019automatic,DroneScan20}. For many downstream applications, abstracted scene proxies that approximate this dense geometry~\cite{kelly2017bigsur,kamra2022lightweight} are often preferred due to their efficiency in transmission, real-time rendering, and other practical uses.

\begin{figure}
	\includegraphics[width=.95\linewidth]{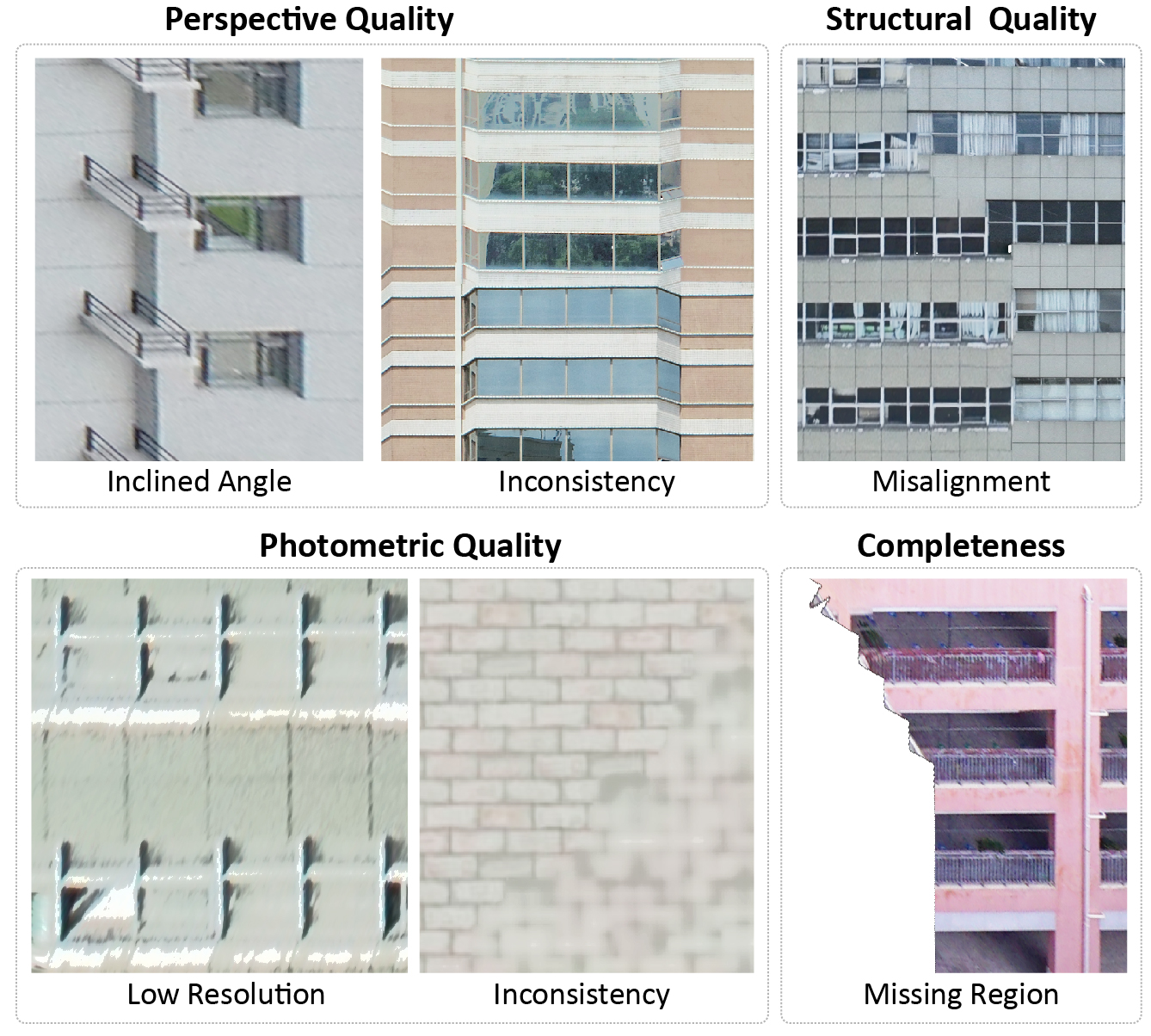}
	\caption{The texture quality evaluation system consists of four aspects: Perspective quality, photometric quality, structural quality and completeness. In each aspects, there remains visual artifacts in the texture maps of the current reconstructed architectural proxies.}
	\label{fig:artifacts}
\end{figure}

A typical pipeline for generating realistic proxy buildings involves several key stages: image acquisition, dense mesh reconstruction, proxy abstraction, and texture mapping~\cite{Kelly2018FrankenGAN,TwinTex23}. Among these, the initial stage, image acquisition, often defines the upper bounds for the quality of the final output. However, to the best of our knowledge, current UAV-based urban reconstruction approaches primarily emphasize capturing detailed scene geometry~\cite{smith2018aerial,DroneScan20}.
None of the existing methods explicitly keep the quality of final texture maps in mind during path planning. 
This oversight often leads to noticeable visual artifacts, especially after complex color remapping procedures in architectural proxy models, as shown in Fig.~\ref{fig:artifacts}.

A straightforward solution is to separately collect photos targeting high-quality texture maps and detailed geometry.
However, in practice, there often exist differences on the calibrated coordinate system or camera settings between these two rounds of image collection.
These inconsistencies can lead to significant misalignment between the reconstructed geometry and textures. Addressing such misalignments is typically labor-intensive, time-consuming, and highly cumbersome.

In this paper, we introduce the urban \textit{geometry and texture co-capture} problem, which focuses on collecting images in a single flight to reconstruct both high-quality texture maps and detailed geometry.
This joint capture task presents significant challenges due to fundamental differences in the requirements for geometry and texture acquisition, primarily stemming from:
i) \textit{Angular disparity}. Geometry reconstruction requires photo disparity in viewing directions, while such angular variation introduces perspective inconsistencies that degrade texture fidelity.
ii) \textit{Coverage requirements}. High-quality texture mapping often requires only a single well-framed view per surface region, whereas geometry reconstruction depends on denser multi-view coverage. Texturing facades with multiple inconsistent photos, on the contrary, would reduce texture quality.
iii) \textit{Shooting time}. While the order in which images are captured has little impact on geometry, it significantly affects the photometric quality of the textures. Non-continuous capture can result in inconsistent lighting conditions, such as varying shadows or sunlight, which manifest as visual artifacts in the stitched texture maps.
Therefore, capturing imagery suitable for both geometry and texture reconstruction demands a carefully balanced drone trajectory, optimizing the number of photos, their spatial positions, viewing angles, and capture sequence.

Our primary goal is to jointly maximize the quality of the scene geometry and texture reconstructed using RGB images captured along the UAV trajectory. We also enforce safety constraints, such as minimum/maximum allowable altitude and safe distances from architectures. Finally, we incorporate flight efficiency by considering the travel cost associated with the planned UAV path. A series of experiments are conducted to validate our approach.
In summary, our work makes the following contributions:
\begin{itemize}
	\item We pose the \textit{geometry and texture co-capture} problem, and propose a novel framework for reconstructing textured structural models using only a 2D map of a complex urban scene and a predefined safe flight altitude.
	
	\item We present a quality assessment system to measure the facade texture map given views. Given a 2D map and a set of 2D views, we introduce a new metric to quantify the reconstructed texture quality of a building facade in a complex scene, as well as a metric for assessing the contribution of each view to the facade's texture quality.

	\item We present multi-objective view planning algorithms designed to efficiently capture compact photos for the reconstruction of high-quality proxy models and texture maps.
	
\end{itemize}

\section{Related Work}\label{sec:rw}

\subsection{Reconstruction-oriented Aerial Images Collection}

The field of drone photogrammetry is rapidly evolving, allows for the acquisition of three-dimensional information about target objects from captured image data.
\citet{smith2018aerial} devised a reconstruction feasibility heuristic method to ensure the camera positions and directions are sufficiently good for multi-view reconstruction. Based on this heuristic-defined metric, regularly generated candidate views can be continuously adjusted to optimal positions and directions.
\citet{koch2019automatic} aimed at detailed and comprehensive small-scale reconstruction, utilizing neural networks to extract semantic information of target objects and optimizing the flight path of objectives through discrete optimization using submodularity and photogrammetric heuristics. \citet{DroneScan20} proposed an adaptive drone path planning algorithm, that utilizes 2D maps and satellite images of the target area. They compute a rough 2.5D model for the scene based on the relationship between buildings and corresponding shadows. Then, they use a Max-Min optimization method to select the minimum viewpoint set and maximize reconstruction quality while maintaining the same number of viewpoints.

To balance reconstruction quality and cost, it is essential to design streamlined drone routes \cite{zhou2020survey, zhang2021continuous}. Model-free methods lack any prior information about the target structure or scene, making computing optimal paths challenging. Model-free methods are mainly categorized into boundary-based planning \cite{dai2020fast, batinovic2021multi, meng2017two,liu2021aerial}, volume-based planning \cite{batinovic2022shadowcasting, song2020online, wang2020efficient}, and surface-based planning \cite{chen2005vision, schmid2020efficient, hardouin2020surface}. Boundary-based planning is one of the most widely used exploration methods for mobile robots. \citet{dai2020fast} achieved rapid and efficient exploration performance by tightly integrating octree-based occupancy mapping, boundary extraction, and motion planning. \citet{batinovic2021multi} proposed a 3D exploration planner based on boundary detection (between explored and unexplored areas), consisting of algorithms for detecting boundary points, clustering them, and selecting the optimal boundary points to explore.
\citet{schmid2020efficient} proposed a detection planning method for reconstructing surface models, using a new RRT method to continuously expand a single tree to achieve maximum utility for global coverage. 
Several commercial flight planning software tools, such as Pix4Dcapture~\footnote{https://www.pix4d.com/product/pix4dcapture/}, DJI Terra~\footnote{https://enterprise.dji.com/cn/dji-terra}, and DroneDeploy~\footnote{https://www.dronedeploy.com/}, support automated flight planning for large-scale photogrammetry and 3D scene reconstruction.

Existing image collection efforts aim to reconstruct more complete and accurate object geometry, considering factors such as energy consumption, collision threats, and path length, while maximizing reconstruction quality \cite{liu2021aerial,liu2022learning,song2021view}.
However, current efforts have not addressed how to capture high-quality image data for the reconstruction of high-quality texture maps for structured proxy models, let alone under limited information. With the large amount of images captured, it remains difficult to achieve texture maps with high level of visual quality.

\begin{figure*}
	\centering
	\includegraphics[width=\linewidth]{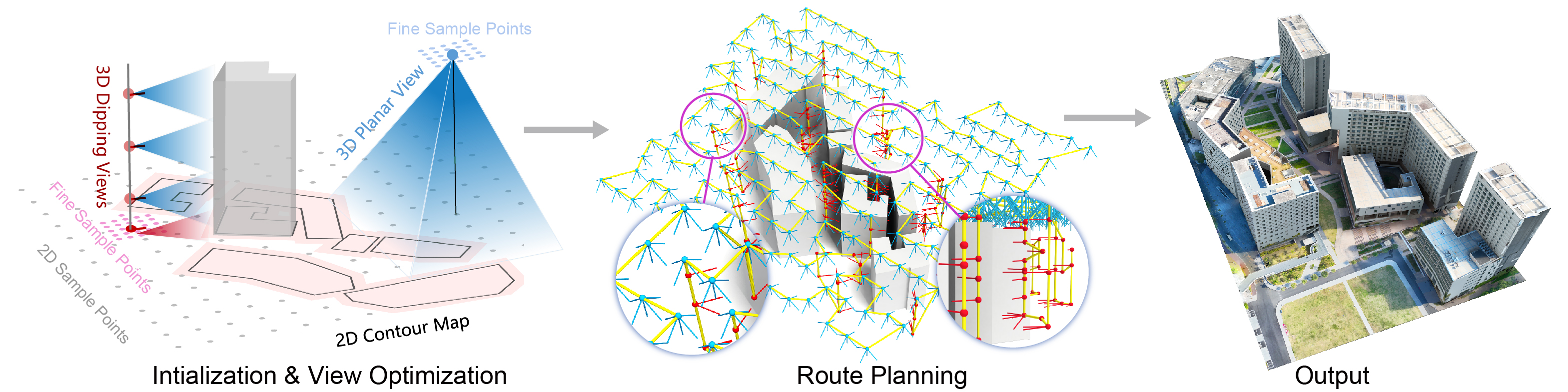}
	\caption{The overview of our framework. The blue cameras represent planar views, while the red cameras correspond to dipping views.}
	\label{fig:overview}
\end{figure*}

\subsection{Structured Geometry and Texture Reconstruction}

Creating structured models from raw data has become an increasing demand in urban digitization.
Such proxy models are generated manually or by procedural modeling~\cite{sinha2008interactive} with arbitrary topology. However, such approaches are labor-intensive and often require expert knowledge to define the model parameters. Existing automated methods typically abstract parameterized primitives from dense mesh and assemble them into polygonal scenes~\cite{garland1997surface, Monszpart2015rapter, salinas2015structure, Nan2017polyfit, kelly2017bigsur, fang2020connect, bouzas2020structure, Pan2022Efficient, Guo2022}.

Compared with a large number of structured scene geometric reconstruction methods, texturing such proxy scenes has been less explored. There are only several methods that perform the generation of texture maps together with a lightweight geometric reconstruction~\cite{coorg1999extracting,sinha2008interactive, garcia2013automatic, huang20173dlite, maier2017intrinsic3d, wang2018plane}.
~\cite{wang2002recovering} proposes an iterative weighted-average algorithm to recover high-quality consensus facade textures, assuming largely planar wall surfaces and predominantly rectangular micro-structures. 
Recently, data-driven methods have been proposed to synthesize texture details for coarse meshes of urban areas~\cite{Kelly2018FrankenGAN, georgiou2021projective}. Although generating various styles of textures, these approaches demonstrated heavy reliance on the generative model trained on a specific dataset, and the synthesized textures lack realism compared to the original real scenes.
Recent NeRF-based methods~\cite{mueller2022instant} and 3D Gaussian Splatting~\cite{kerbl20233d} have attracted significant attention for reconstructing and rendering complex scenes~\cite{lin2024vastgaussian, huang20242d}, achieving both realistic novel views and high-quality 3D shapes and textures~\cite{metzer2022latent,baatz2022nerf}. However, when processing very high-resolution input photos, these methods demand extremely large memory, which limits their applicability in our scenario.

Most closely related to our work is TwinTex~\cite{TwinTex23}, which provides an applicable framework for texturing architectural proxy with high quality given highly uneven camera distribution.
However, input photos with extremely limited quality still result in obvious artifacts.
Therefore, the quality of source images plays a crucial role in the quality of reconstructed texture maps, even for well-designed texture reconstruction methods.

\subsection{Image-based Texture Map Generation}

Aiming for realistic 3D reconstruction, high-quality texture mapping onto the reconstructed geometry has been extensively studied, where a large volume of methods are developed to optimize color textures.
Previous image-based texture mapping methods can be classified into three categories: (1)~\textsl{blending-based} methods select multiple images for each face and blend them into textures by using different weighting strategies~\cite{bernardini2001high, callieri2008masked,lee2020texturefusion}, (2)~\textsl{projection-based} methods associates each face or vertex with one appropriate input image for creating a texture patch, and then conduct seam optimization to avoid visible seams between adjacent patches~\cite{Debevec1996Modeling,lempitsky2007seamless,wang2018seamless,gal2010seamless, waechter2014let,fu2018texture,fu2020joint}, (3)~\textsl{warping-based} methods jointly rectify the camera poses and geometric errors, which usually use local image warping~\cite{zhou2014color} or patch-based optimization~\cite{bi2017patch} to avoid texture misalignment caused by geometric error and camera drifting.
Although high-quality texture maps can be generated from the above methods, the resultant quality of texture mapping largely depends on the quality of collected photos.

\section{Overview}
\label{sec:overview}

Our input consists of a 2D map in which buildings within the target reconstruction area are recognized as poly-line contours, see Fig.~\ref{fig:teaser} and Fig.~\ref{fig:overview}.
The precise heights of these buildings are unknown, and only an altitude $H$ is provided to ensure the UAV can fly safely above all structures. 
Our goal is to generate a safe aerial path for the UAV to co-capture images for the reconstruction of both structured geometry and high-quality texture maps at low cost. To the best of our knowledge, this task remains unexplored and is particularly challenging due to the limited site-specific information.

The pipeline of our method is illustrated in Fig.~\ref{fig:overview}.
The core idea of our coarse-to-fine aerial path planning is to generate a set of dipping views (red views in Fig.~\ref{fig:overview}) that maximize facade texture quality (Sec.~\ref{sec:views-dipping}), alongside a set of planar views (blue views in Fig.~\ref{fig:overview} locating at a fixed altitude) aimed at maximizing ground and roof texture quality, as well as reconstructed geometry quality (Sec.~\ref{sec:views-planar}).
Based on these 3D viewpoints, we then construct a structure-aware aerial path, which guides the UAV to capture the necessary images. These images are ultimately used to reconstruct both a proxy model and high-quality texture maps (Sec.\ref{sec:reconstruction}). 

\section{Dipping Views}
\label{sec:views-dipping}

Our input 2D scene map consists of a set of poly-lines $\{ f_i \}$, where each 2D line segment corresponds to a 3D building facade.
In this section, we aim to generate a set of 3D dipping views $\{V=(P, S)\}$ where each view $V$ is defined by a position $P$ and a viewing direction $S$. These views are intended to capture images that enable the reconstruction of high-quality textures for each building facade.

While previous geometry-oriented methods promote variation in viewing directions~\cite{zhou2020survey}, we adopt a texture-oriented strategy in which the UAV vertically descends from a safe altitude $H$, capturing a sequence of images along the trajectory with a consistent viewing direction.
The problem is then simplified into the generation of 2D dipping views $\{ v=(p, s) \}$ on the 2D map, where $p$ denotes the 2D dipping position and $s$ denotes the 2D viewing direction.

This section begins with the definition of a texture quality assessment system, followed by two metrics measuring facade texture quality. Guided by the metrics, a greedy algorithm is developed to initialize a few 2D dipping views. Then, we adjust both the position and viewing direction of 2D dipping views to maximize the quality of the final reconstructed facade texture maps, while reducing the capturing cost over the entire scene. Finally, we lift each 2D dipping view into a 3D view sequence. 
Note that photos are captured only at 3D views, not at 2D views.

\subsection{Quality Metrics}
\label{sec:quality}

We begin by proposing a novel quality system is to comprehensively measure the texture quality of a facade $f_i$ given a set of 2D views, $\{ v \}_{f_i}$, even before obtaining any texture maps:

\begin{itemize}[leftmargin=*]
\item \textit{Perspective quality} $Q_s$ assesses the viewing direction consistency and frontality of $\{ v \}_{f_i}$.

\item \textit{Photometric quality} ${Q}_{d}$ assesses the sharpness and resolution consistency of the resulting texture map. Since the distance between a view and the target facade directly affects image resolution, we favor views with small and similar distances to the target facade.

\item \textit{Structural quality} ${Q}_{u}$, measures the structural coherence of the facade in the resulting texture map. Due to potential calibration errors, incorporating a larger number of photos during texture remapping can lead to increased structural distortions (\eg distorted linear features, misaligned global structures, etc). Hence, we encourage less number of photos in $\{ v \}_{f_i}$, resulting in lower texel coverage.

\item \textit{Completeness} ${Q}_{c}$ measures the ratio of facade $f_i$ that can be observed by $\{ v \}_{f_i}$.
\end{itemize}

The detailed formulations of the quality terms are provided in the supplementary document. Following the above quality system, we introduce two metrics to guide the entire planning process.

\paragraph{Facade quality} We first define the texture quality metric for facade $f_i$ with respect to $\{ v \}_{f_i}$. The quality $Q(\{ v \}_{f_i}, f_i)$ is computed as the combination of four quality terms:
\begin{align}
	\label{eq:facade_quality}
	& Q(\{ v \}_{f_i},f_i) = Q_{s}(\{ v \}_{f_i},f_i) + Q_{d}(\{ v \}_{f_i},f_i) \\ \nonumber
	& \qquad \qquad \quad + Q_u(\{ v \}_{f_i}, f_i) + Q_c(\{ v \}_{f_i}, f_i).
\end{align}

\paragraph{View-Facade quality} Next, we introduce a metric to evaluate the contribution of an individual view $v_j$ in $\{ v \}_{f_i}$ to the overall texture quality of facade $f_i$. We define the quality ${Q}(v_j,f_i)$ as:
\begin{align}
	\label{eq:view_quality}
	& {Q}(v_j,f_i) = \lambda_1 {Q}_{s}(v_j,f_i) + \lambda_2 {Q}_{d}(v_j,f_i) \\ \nonumber
	& \qquad \quad \quad + \lambda_3 {Q}_u(v_j,f_i) + \lambda_4 {Q}_c(v_j,f_i),
\end{align}
where $\lambda_1=0.1$, $\lambda_2=0.85$, $\lambda_3=0.3$, and $\lambda_4=0.1$ are the balancing coefficients.

\subsection{Dipping Views Initialization}
\label{sec:dipping-initialization}

In this subsection, we aims to generate a small set of high-quality 2D dipping views $\{ (p, s) \}$ inside the target 2D region. However, properly determining a few views along with their positions and viewing directions to maximize facade texture quality remains highly challenging even in 2D space, especially in complex urban environments where occlusion and uncertainty are common.

\begin{figure}
	\centering
	\includegraphics[width=\linewidth]{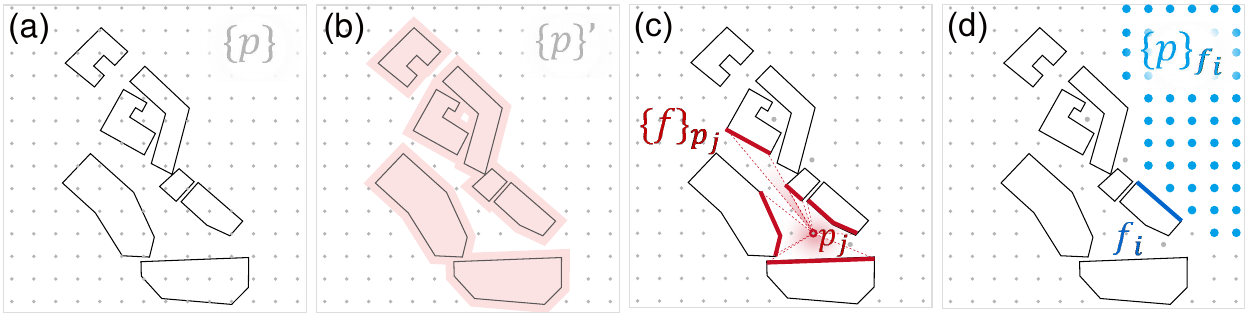}
	\caption{ (a) The 2D regular sample points (grey dots). (b) The red no-dipping zone and \textit{candidate dipping points} after elimination. (c) All the visible facades and corresponding visible regions (marked in red) of a candidate dipping point (red circle). (d) All the candidate dipping points (marked in blue) that can observe the whole/portion of a facade plane marked in blue.}
	\label{fig:visible}
\end{figure}

We discretize the parameter space of view position and viewing direction to find feasible initial views. To start, we generate a dense set of regular 2D sample points $\{ p \}$ over the target 2D region, see the gray points in Fig.~\ref{fig:visible}(a).
To avoid crashing, we eliminate the sample points located inside the no-dipping zone (red regions in Fig.~\ref{fig:visible}(b)) and obtain the subset of \textit{candidate dipping points} $\{ p \}'$.
Currently, each point in  $\{ p \}'$ is associated with no viewing directions, and thus has no limitation on the field of view unless occluded by buildings.
Here, for a point $p_j$, we perform a visibility check to retrieve all its visible facades $\{f\}_{p_j}$ and corresponding visible regions (Fig.~\ref{fig:visible}(c)).
In turn, given a facade $f_i$, we can obtain a set of candidate dipping points $\{ p \}_{f_i}$ in $\{ p \}'$ that can observe the entire or a portion of $f_i$ (Fig.~\ref{fig:visible}(d)).

A straightforward approach is to perform a brute-force search, evaluating all possible combinations of 2D positions and viewing directions to generate feasible views for each facade. However, this method is computationally intensive. To address this, we simplify the problem by decoupling the selection of viewing direction and position, significantly reducing the computational cost.

\paragraph{Viewing direction initialization}
For $\{ p \}_{f_i}$ capturing facade $f_i$, we assign all the points with the same viewing direction $s_i$. With this strategy, for facade $f_i$, the perspective consistency term in $Q_s$ is maximized and simplified into a perspective frontality term.
Hence, to find an initial viewing direction delivering high-quality texture, we search sample viewing directions around the inverse facade normal of $f_i$ with a small step. The sample direction delivering the highest facade quality $Q(\{ v \}_{f_i}, f_i)$ is employed as $s_i$.

\begin{figure}
	\centering
	\includegraphics[width=\linewidth]{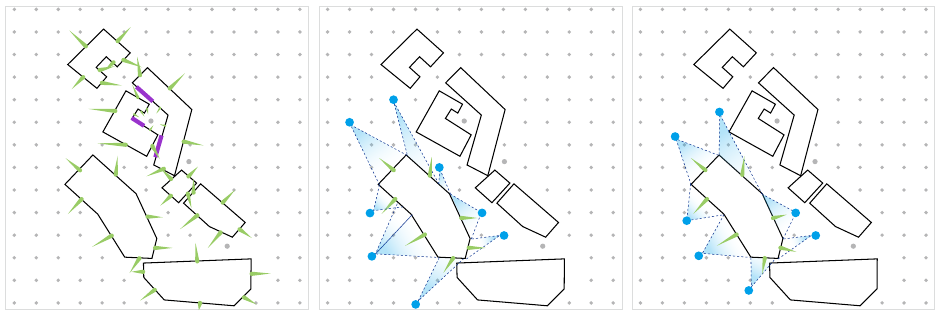}
	\caption{Left: The results of initializing the viewing directions (green vectors) of all the facades in a scene containing five buildings. The unobserved facade regions are visualized in purple. Middle: Results on the selected initial dipping points (blue points) and viewing directions of building facades. Right: Results on the optimized dipping points and viewing directions of building facades.}
	\label{fig:dipping}
\end{figure}

The initialized viewing directions of all the facades in a scene are shown as green vectors at the left in Fig.~\ref{fig:dipping}. As observed, most viewing directions are perpendicular to their facades, except in areas with occlusions or no-dipping zones.

\paragraph{Dipping points selection} With assigned viewing directions, the field of views as well as visible facade regions are then settled. Next, we seek to select a small subset of 2D dipping points $\{ p \}''$ from $\{ p \}'$ to increase facade texture quality and decrease capturing cost. 
One possible approach is to independently select dipping points with high view-facade quality for each facade. However, this overlooks the fact that a dipping point may observe multiple building facades. To address this, we propose an iterative approach to select a small set of points that collectively cover the entire scene.
At iteration $t$, we calculate the quality of $p_j$ with respect to all visible facades $\{ f \}_{p_j}$. Here the 2D view quality $\widetilde{Q}$ of $p_j$ is computed as:
\begin{equation}
	\widetilde{Q}(p_j, \{ f\}_{p_j}) = \sum_{f_i \in \{ f\}_{p_j}} {Q}((p_j,s_i), f_i).
\end{equation}

We iteratively remove the point with the lowest quality $\widetilde{Q}$ in $\{ p \}'$.
This process converges until any point removal will decrease the completeness $Q_c$ of facades.
The blue points in middle of Fig.~\ref{fig:dipping} denote the selected dipping views for a building in the Polytech scene.

\begin{figure}
	\centering
	\includegraphics[width=.6\linewidth]{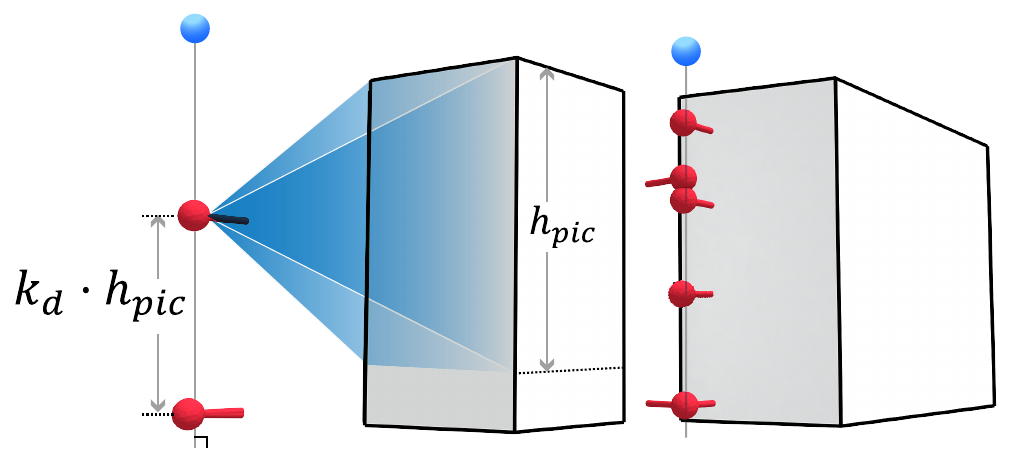}
	\caption{Left: A sequence of 3D views (red) dipping from point (blue) capturing a single facade. Right: Two sequences of 3D views (red) dipping from the same point capturing two different facades.}
	\label{fig:lifting}
\end{figure}

\paragraph{3D dipping views generation}
Finally, for a 2D dipping point $p_j$ capturing target facade $f_i$, a vertical sequence of 3D views will be generated at intervals of $k_d \cdot h_{pic}$ starting from height $H$, denoted as $\{ V \}_{p_j, f_i}$. The viewing direction of all the views will follow $s_i$, as shown in Fig.~\ref{fig:lifting}.
All 3D views along the dipping trajectory maintain a consistent distance to facade $f_i$, thereby maximizing the photometric consistency in $Q_d$.
We let $k_d=0.8$ to ensure texture completeness and enough overlaps among the projected regions of adjacent views for the image stitching in texture remapping.
We define $h_{pic}(V_a,f_i) = h_{sensor} \cdot d(V_a,f_i)/{f}$ as the projected sensor height of view $V_a$ onto plane $f_i$, where $h_{sensor}$ and $f$ is the height and focal length of the utilized camera sensor, respectively.

Note that constrained by the minimal flying height, the facade regions close to the ground may not be covered by any dipping views. We introduce an additional viewing direction at the lowest viewing points, ensuring vertical coverage of the full building facade.

\subsection{Dipping Views Optimization}
\label{sec:optimization}

\paragraph{Flight hovering cost.}
Given 3D views, the UAV needs to slow down and hover to capture a photo at the position $P$ of each view $V$, then speed up heading to the next view.
In practice, allowing multiple views to share the same position can reduce the total number of required hover points, significantly lowering the energy cost of image acquisition. In this work, we quantify flight cost by the number of distinct hovering positions. For instance, the co-located 3D view sequences shown on the right in Fig.~\ref{fig:lifting} have six views and five hovering positions.

For each 2D dipping view $v_j=(p_j, s_i)$ capturing facade $f_i$, we can lift a sequence of 3D views $\{V\}_{p_j,f_i}$.
The hovering cost $C({p_j})$ of a dipping point $p_j$ is defined as the total number of all the 3D hovering positions lifted from $p_j$.
The hovering cost $C$ over the scene can be computed as: $C = \sum_{p_j \in \{ p \}''} C({p_j})$.

In complex scenes, multiple sequences of 3D views are often lifted from the same dipping point to capture different facades, as illustrated in Fig.~\ref{fig:lifting}.
Consider two such lifted views: $V_a$ capturing facade $f_o$, and $V_b$ capturing facade $f_m$, both originating from the same dipping point $p$. If their distance is smaller than the projected overlap $\tau_d = (1-k_d) \cdot min\{h_{pic}(V_a,f_o), h_{pic}(V_b,f_m)\}$, $V_a$ and $V_b$ can be merged into their midpoint. This merged position represents a shared hovering point with dual viewing directions, trading off a small amount of vertical facade coverage to reduce flight cost.
We compute the cost savings $\delta(d)$ resulting from merging $V_a$ and $V_b$ based on their distance, $d(V_a,V_b)=||V_a - V_b||_2$:
\begin{equation}
	\delta(d)= \begin{cases}
		\dfrac{1}{2} Gauss(d), &d \le \tau_d\\
		0, & others
	\end{cases}
\end{equation}
where $Gauss(\cdot)$ is the gaussian function with $\mu=0$ and $\sigma=\tau_d/3$. This function tend to merge closer views, while the view pair with distance larger than the threshold $\tau_d$ will not be merged.

Then, the hovering cost of $\{ p \}''$ capturing all the facades can be formulated as:
\begin{equation}
	C = \sum_{p_j \in \{ p \}''} \left( C({p_j})- \sum_{V_a } \sum_{ V_b } \delta(d(V_a, V_b)) \right),
\end{equation}
where $V_a$ and $V_b$ are from two different view sequences dipping from $p_j$.

\paragraph{Multi-objective function.}
Given the initialized views, we seek to further reduce the hovering cost $C$ while also enhancing facade quality. This problem is extremely challenging due to the complexity of the scene and limited information. Instead of directly adjusting 3D views which can easily affect the texture quality and are difficult to control, we propose to fine-tune the 2D dipping views within a constrained range and subsequently lift them to 3D.
The variables are all the 2D points position and 2D viewing directions.
This problem is formulated into a multi-objective optimization. We seek to minimize the following objective function vector $\boldsymbol{{G}}(v)$ of each 2D dipping point, observing facades $\{ f_i, \dots, f_n \}$:
\begin{equation}
	\boldsymbol{{G}} =\left( C, -Q(\{v\}_{f_i}, f_i), \dots , -Q(\{v\}_{f_n}, f_n)
	\right).
	\label{eq:multi-objectives}
\end{equation}

The dominance relationship between function vectors $\boldsymbol{G}$ and $\boldsymbol{G^{'}}$ with $(g_1, \dots, g_k)$ is defined as $\boldsymbol{G} \prec \boldsymbol{G^{'}}$, if and only if $\forall \, i \in (1, \dots, k)$: $g_i\le g_i^{'}$ and $\exists \, i \in (1, \dots, k)$: $g_i<g_i{'}$~\cite{yang2013multiobjective}.
This is an NP problem which is hard to solve.

We propose an approach to iteratively optimize the 2D dipping views one at a time. In each iteration, three operations are performed sequentially:
i) \textit{Dipping point adjustment}.
For each dipping point $p$ in  $\{p\}''$, we sample its surrounding space with step $\tau_p$ (the pink points in Fig.~\ref{fig:overview} on the ground). 
The sampled point with the smallest $\boldsymbol{G^*}$ is updated as $p$. 
ii) \textit{Facade direction adjustment}. For the viewing direction $s_i$ of a facade $f_i$, we sample its neighboring space with step $\tau_s$.
The sampling direction with the smallest $\boldsymbol{G^*}$ is selected as the new viewing direction of $f_i$ and $\{ v \}_{f_i}$. 
iii) \textit{Dipping view removal}. The redundant dipping views with no contribution to any facade will be removed.
This algorithm converges when all the dipping views remain unchanged.
Our extensive experiments show that $5$ iterations are often sufficient for convergence.

The optimized result on a single building in the Polytech example is shown on the right in Fig.~\ref{fig:dipping}. We can observe that the total number of dipping views is reduced, the adjusted dipping points are closer to its facade and the adjusted viewing directions are more perpendicular to the facade.
These optimized 2D dipping views will be lifted to 3D views and remain fixed in later steps, as the red views shown in Fig.~\ref{fig:scene-sampling}.

\section{Planar Views}
\label{sec:views-planar}

\begin{figure}
	\centering
	\includegraphics[width=\linewidth]{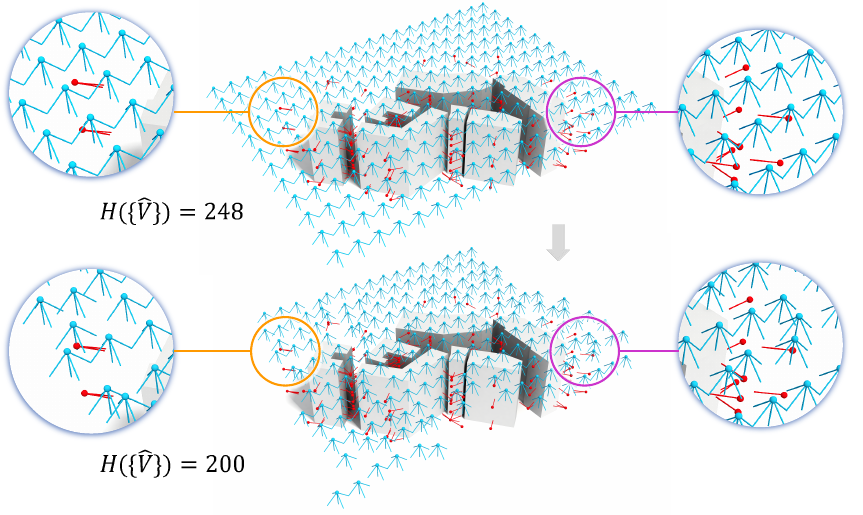}
	\caption{Top: The 3D views before optimizing the planar views (blue views), Bottom: The 3D views after optimization. The red dipping views remain unchanged.
	}
	\label{fig:scene-sampling}
\end{figure}

Currently, we have sequences of 3D dipping views $\{(P, S)\}$ capturing high-quality facade material with low cost. However, photos captured from these dipping views are insufficient for reconstructing the scene geometry due to a lack of viewing direction variations and multiple-view coverage.
In this section, we aim to generate another set of 3D views $\{\widehat{V}=(\widehat{P}, \widehat{S})\}$ capable of further reconstructing the scene geometry, as well as textures of the roof and ground.

To begin, we lift all regular 2D sample points $\{p\}$ to altitude $H$, and let UAV capture five photos at each position: one with a vertical viewing direction and four with tilted perspectives (blue view in Fig.~\ref{fig:scene-sampling} top). This strategy is energy-efficient and has been shown to be sufficient for accurate scene geometry reconstruction~\cite{DroneScan20}.
Additionally, these vertical views result in texture maps for the roof and ground with extremely high perspective quality and photometric quality. Consequently, the texture quality of the ground plane can be simplified to $Q=Q_u+Q_c$.

Next, we seek to reduce the number of hovering positions in $\{\widehat{V}\}$ without compromising geometry and texture quality. We denote the number of 3D hovering positions in views set $\{ \widehat{V} \}$ as $H(\{ \widehat{V} \})$. Unfortunately, any modifications to these 3D planar views simultaneously impact both geometry and texture quality, making the refinement of views particularly challenging.
To evaluate the quality of $\{\widehat{V}\}$, we generate a 2.5D model according to the input, then sample the model surface into dense sampling points.

For geometry reconstruction, following~\citet{DroneScan20}, we compute the reconstructability $Q_r$ of all the surface sample points, and compute the redundancy $r(\widehat{V})$ of each planar position as the sum of redundancy on five co-located views. 
For texture reconstruction, we calculate the facade quality $Q(\{\widehat{V}\}', f_g)$ of the ground and roof plane $f_g$ only using the vertical views $\{\widehat{V}\}'$ at each 3D planar position $\widehat{P}$.

\paragraph{Multi-objective function.}
To sum up, our target is to simultaneously improve the quality of reconstructed texture maps $Q$, reconstructed scene geometry $Q_r$, while reducing the flight hovering cost $H(\{\widehat{V}\})$ via adjusting the hovering positions of $\{\widehat{V}\}$. We formulate this problem as a multi-objective optimization, i.e., minimizing an objective function vector $\mathbf{Y}(\widehat{V})$:
\begin{equation}
	\mathbf{Y}(\widehat{V})=\left( -Q(\{\widehat{V}\}', f_g), -Q_r, H(\{\widehat{V}\}) \right).
\end{equation}

This is an NP-hard problem, making it extremely challenging to find an optimal solution. Here, we introduce a greedy heuristic to progressively minimizing our objectives via refining the 3D planar views.
Each iteration involves two operators:
i) \textit{View position adjustment}, where the neighboring planar space of $\widehat{P}_j$ is searched with a step size $\tau_p$ (the light blue points in Fig.~\ref{fig:overview} over the sky). The position $\widehat{P}^*$ yielding the smallest $\boldsymbol{Y^*}$ is selected as the new position of $\widehat{P}_j$.
ii) \textit{View removal}, where we greedily and globally remove the most redundant planar views, $\widehat{V}_j=\arg \max_j r(\{\widehat{V}\})$, one by one until any removal will generate uncovered region or a surface sampling point whose reconstructability is smaller than a predefined threshold {$\tau_r$}.
The algorithm converges when all the planar views remain unchanged.

The views before and after optimization for the Polytech example is shown on the bottom in Fig.~\ref{fig:scene-sampling}. We can observe that the total number of planar views (blue views) is reduced, and the density of adjusted views is larger around buildings.

\section{Scene Capture and Reconstruction}
\label{sec:reconstruction}

\paragraph{Path planning}
After determining the dipping views $\{ (P, S) \}$ and planar views $\{ (\widehat{P}, \widehat{S}) \}$ from which the drone needs to capture images, we connect them to form a continuous path. We model the aerial path planning as a standard Traveling Salesman Problem (TSP). In this context, each view is represented as a node in the graph. We construct a fully connected graph and find the shortest path within the safe zone for all view pairs~\cite{alt1988visibility}. Note that if two neighboring views are not captured consecutively, the ambient illumination may vary significantly between shots, leading to high photometric inconsistency (such as shadows, sunlight color variations, etc.).
Considering the changes in orientation and the topological relationships of target facades between a view pair $( V_i, V_j )$, the cost function for an edge connecting them is defined as:
\begin{equation}
	e(V_i, V_j) = w_p l(P_i,P_j) exp( \dfrac{\alpha}{l(P_i,P_j)} ),
\end{equation}
where $l(P_i,P_j)$ is the shortest straight-line distance from position $P_i$ to $P_j$ in the safe flight space. $\alpha$ is the angle between $V_i$ and the line of sight direction of $V_j$. $w_p$ is the topology co-efficient to ensure that the views capturing the same facade and adjacent facades are taken within a short period.
In practice, we set $w_p=0.5$ when they are targeting the same plane, $w_p=0.75$ for adjacent planes, and $w_p=1$ in other cases. A minimal cost path in this graph corresponds to an alternating sequence of views that can be efficiently found by solving the TSP problem~\cite{helsgaun2015solving}.

\paragraph{Abstracted geometry and texture reconstruction.}
Using the planned aerial path, we collect RGB photos with a UAV. {The captured images are calibrated with the commercial software ContextCapture~\footnote{https://www.soarscape.com/} to reconstruct a high-precision 3D model. The ContextCapture follows the standard 3D reconstruction pipeline, including: feature mapping, structure from motion (SfM), multi-view stereo (MVS), and surface reconstruction.}
Next, the dense model is input into a structure-aware reconstruction algorithm~\cite{bauchet2020kinetic} to create an abstracted proxy. Using the calibrated photos and the abstracted proxy, we then employ TwinTex~\cite{TwinTex23} to generate high-quality texture maps, producing the final textured scene.
With our framework, the calibrated photos, dense models, and abstracted models are all aligned within the same coordinate system. This integration allows users to generate high-quality textured scenes in a single workflow, eliminating the need for multiple rounds of photo acquisition, calibration, and manual alignment between different coordinate systems.

\section{Experimental Results}
\label{sec:results}

\paragraph{Implementation}
Our method is implemented in C++.
All the presented experimental results are obtained on a desktop computer equipped with an Intel i9-10900k processor with 3.0 GHz and 128 GB RAM.
We ran our algorithm under a fixed set of parameters ($focal \_ length=12.67mm$, $d_{SGD}=4cm$), $overlap \_ ratio=(0.8,0.8)$.
We use $d_{max}$ as the maximum viewing distance allowed for the equipped camera to capture facade photos with acceptable resolution, and $d_{min}$ as the minimum safe distance for a drone to capture facade photos.
We set $d_{min} = 10 \mathrm{m}$ for all scenes. $H$ is defined as the height of the tallest building plus $d_{min}$. Each building footprint is dilated by a radius of $d_{min}$ on the 2D map to define the \textit{no-dipping zone} for the drone (see red regions in Fig.~\ref{fig:visible}(b)).
{A safe flight indicates that all the building components in reality (e.g., bridges, balconies, or holes) should remain within the calculated \textit{no-dipping zone}. We assume the input 2D map provides sufficient detail on building components to ensure flight safety.}

We compare our reconstructed results against oblique photograph (OP)~\footnote{http://www.moutong.net/} and DroneScan (DS)~\cite{DroneScan20}.
The path planning and data capture parameters (\eg $overlap \_ ratio$, $extended \_width$, $H$) of OP are identical to ours. To ensure a fair comparison, we provided 2.5D models with accurate maximum heights as input to DS. The proxy geometry and texture reconstruction process for all the methods follows the identical pipeline described in Sec.~\ref{sec:reconstruction}.

\paragraph{Dataset} To evaluate the proposed planning and reconstruction framework, we conduct a series of experiments on a dataset  with varying building complexity, styles, functions and densities. The dataset includes four real-world scenes (Polytech, SEng, Hitech, ArtSci), three publicly available virtual scenes (Concave, C-1, Bridge-1)~\cite{smith2018aerial,DroneScan20}, and four newly virtual scenes created by an in-field modeler (SN-1, SN-2, Asia-1, Asia-2). Please refer to the supplementary material for more details and self-evaluation experiments.

\begin{table}
	\caption{Comparison of battery consumption under identical flight paths and number of captured views, but with different numbers of hovering positions.}
	\label{tab:hovering-cost}
	\begin{center}
		\resizebox{\linewidth}{!}{
			\begin{tabular}{c c c c c c}
				\midrule[1pt]
				Length (m) & $\#$ Images & Flight ID & $\#$ Hover  & Energy (\%) & Time (min) \\  \hline
				\multirow{2}{*}{895}   & \multirow{2}{*}{250}  &0  & 250    & 86 & 29   \\
				\cmidrule(lr){3-6}
				&   &1   & 119   & 57 & 16\\
				\midrule[1pt]
			\end{tabular}
		}
	\end{center}
\end{table}

\begin{figure*}
	\centering
	\includegraphics[width=\linewidth]{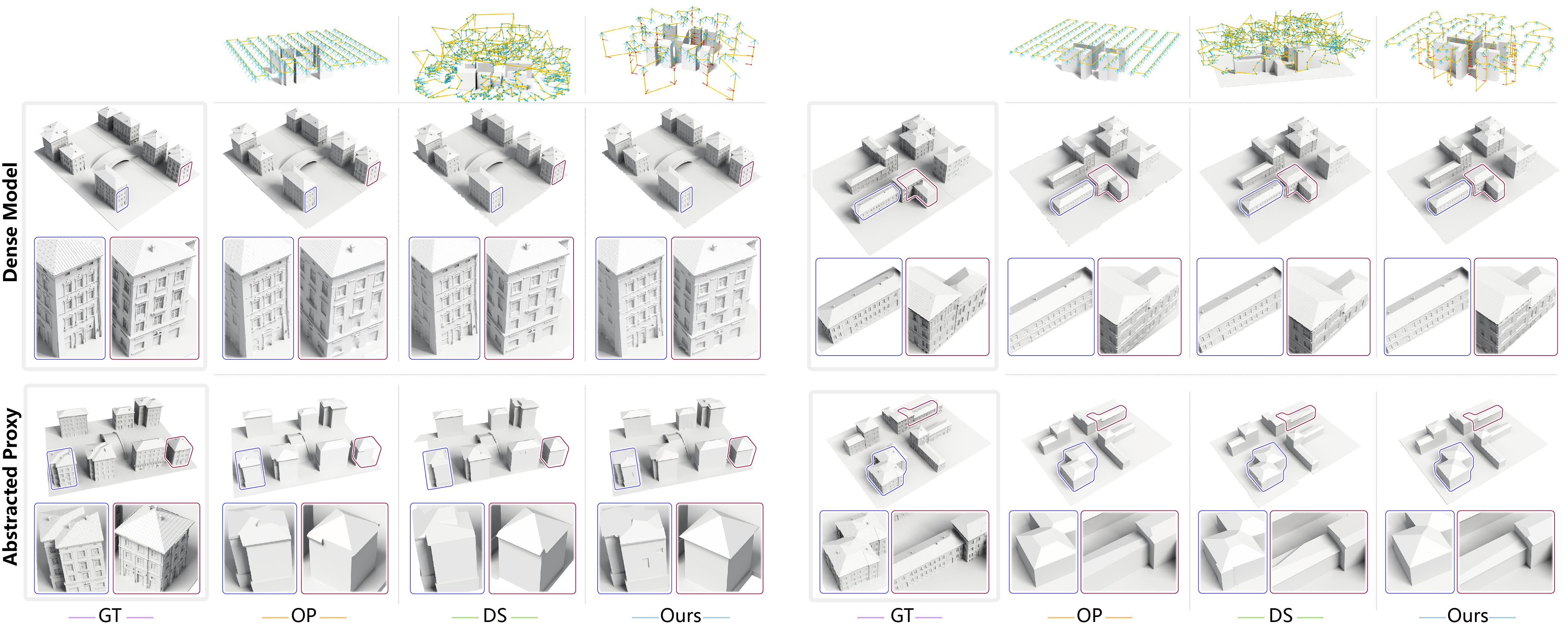}
	\caption{The overall views of the stage results on the reconstructed dense geometry and abstracted geometry and texture maps of two virtual scenes.}
	\label{fig:stage-results-geometry}
\end{figure*}

\begin{table*}
	\caption{Statistics on the capturing and reconstructed geometry results of virtual scenes. $\#$ Hover denotes the total number of hovering positions.}
	\label{tab:quality-stage}
	\begin{center}
			\begin{tabular}{c c c c c c c c c}
				\midrule[1pt]
				\multirow{2}{*}{Scene} & \multirow{2}{*}{Method} & \multicolumn{3}{c}{Capturing} & \multicolumn{2}{c}{Dense Model} & \multicolumn{2}{c}{Proxy Model} \\
				\cmidrule(lr){3-5} \cmidrule(lr){6-7} \cmidrule(lr){8-9}
				&   & $\#$ Images & $\#$ Hover & Trajectory  & Error $\downarrow$ & Comp (\%) $\uparrow$ &Error $\downarrow$ & Comp (\%) $\uparrow$  \\
				\cmidrule(lr){1-1} \cmidrule(lr){2-2} \cmidrule(lr){3-5}  \cmidrule(lr){6-7} \cmidrule(lr){8-9}
				\multirow{3}{*}{C-1}        & OP   & 840   & 168 & \numprint{4558}    &0.76 &88.73 &1.37 & 81.23   \\
				\multicolumn{1}{c}{}        & DS   & 789   & 789 & \numprint{10638}   &0.30 & \textbf{97.07} & \textbf{0.97} & 83.92    \\
				\multicolumn{1}{c}{}        & Ours & 846 & 225 & \numprint{5881}      &\textbf{0.19} &96.59 &1.05 & \textbf{85.56}   \\
				\cmidrule(lr){1-1} \cmidrule(lr){2-2} \cmidrule(lr){3-5}  \cmidrule(lr){6-7} \cmidrule(lr){8-9}
				\multirow{3}{*}{Bridge-1}   & OP    & 720    & 144 & \numprint{3008}  &0.50 &86.98 &1.12 &\textbf{89.84}  \\
				\multicolumn{1}{c}{}        & DS    & 437   & 437 & \numprint{5468}   &0.20 &\textbf{93.26} &\textbf{0.74} &88.07  \\
				\multicolumn{1}{c}{}        & Ours  & 695  & 183 & \numprint{3277}    &\textbf{0.16} &91.97 &0.83 &87.18  \\
				\midrule[1pt]
			\end{tabular}
	\end{center}
\end{table*}

\paragraph{Energy cost and hovering position number.}
To reduce energy consumption, \citet{roberts2017submodular} minimizes the length of trajectory, while \citet{DroneScan20} also reduce views number. Whereas, in this paper, we also consider the impact of number on hovering positions. To evaluate the benefits of reducing hovering positions, we examine a real-world scene and generate flight missions with identical path lengths and numbers of views, but with different numbers of hovering positions. Table~\ref{tab:hovering-cost} reports the energy consumption associated with each mission. Notably, doubling the number of hovering positions results in an 81\% increase in flight time and a 51\% increase in battery consumption. These results indicate that reducing the number of views not only indirectly shorten the flight path but also directly reduces overall energy consumption.

\subsection{Comparison on Stage Results}

We first evaluate the stage results of geometry on two virtual scenes, C-1 and Bridge-1.
The visualization on the aerial flight path, the dense reconstructed model, and the abstracted proxy model, are shown in Fig.~\ref{fig:stage-results-geometry}. 
Following~\citet{smith2018aerial} and \citet{DroneScan20}, we compute two metrics to assess the quality of the reconstructed geometry $M$: i) \textit{Error}, which quantifies how closely $M$ approximates the ground truth model $M^*$; and ii) \textit{Completeness}, which measures the coverage of $M$ relative to $M^*$. These metrics are computed for all mesh vertices. We further analyze the distributions of these values and report the percentage of vertices with values smaller than 95\% in Table~\ref{tab:quality-stage}.

All the statistics are provided in Table.~\ref{tab:quality-stage}. For images capturing, DS tends to generate much longer flight trajectory, even targeting at similar or less number of images. 
Since our method and DS plan views at lower altitudes, the dense models exhibit higher accuracy and completeness for buildings at lower altitude, outperforming the OP method. However, because DS captures more photos per building, it often reconstructs dense geometry with a little higher level of detail than our method. 
In contrast, the abstraction process simplified the fine details (such as windows). Hence, the statistics on proxies shows larger error and smaller completeness compared to the dense models. For a fair comparison on texture maps, we let all the abstracted proxy models share similar level of details~\cite{bauchet2020kinetic}, thus, similar accuracy and completeness.

\begin{figure*}
	\centering
	\includegraphics[width=\linewidth]{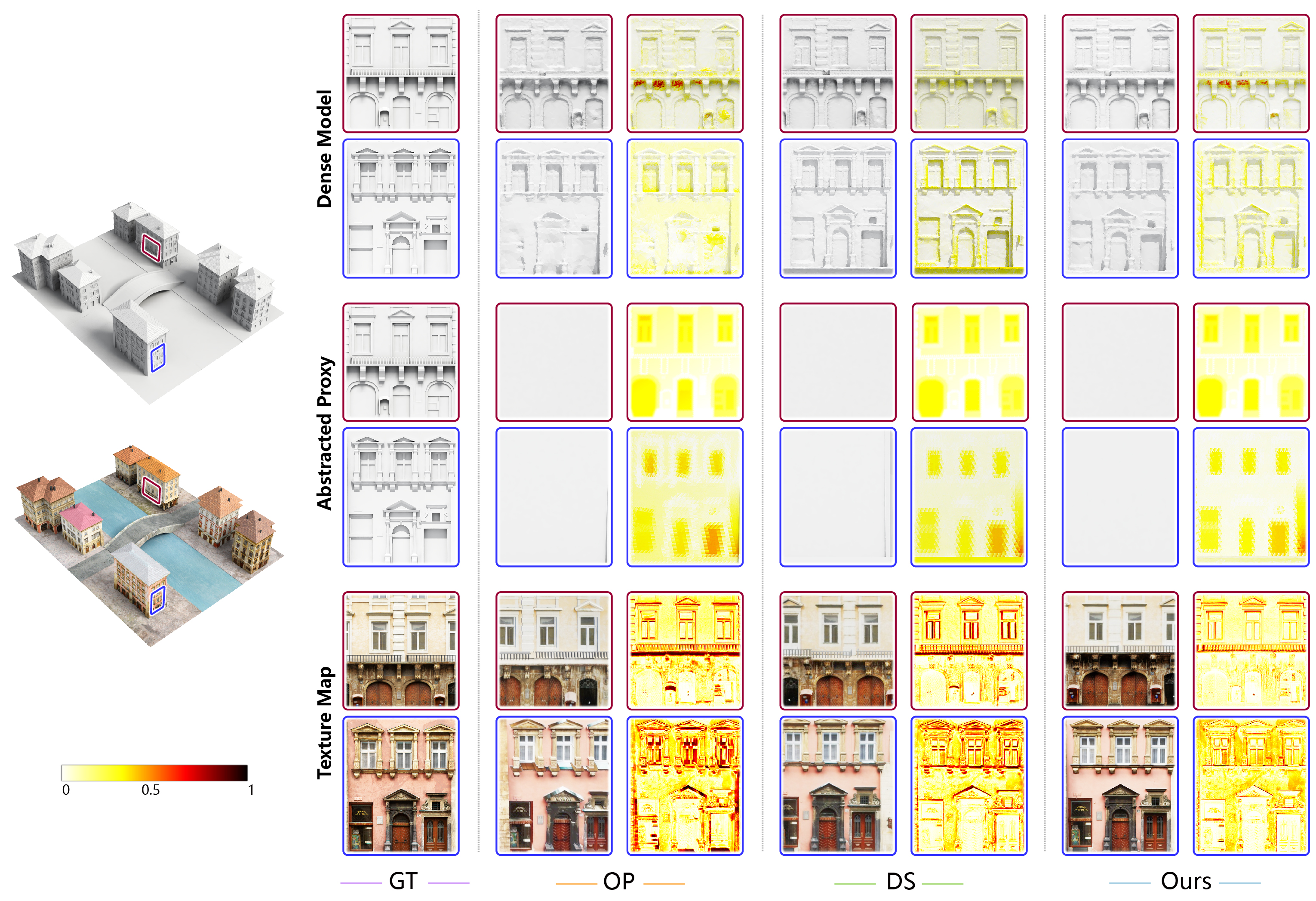}
	\caption{The stage results on the reconstructed dense geometry, abstracted geometry and texture maps of a virtual scene, called Bridge-1. The detailed comparisons are shown in the zoomed-in insets. The color denotes the quality of the reconstructed geometry and texture.}
	\label{fig:stage-results-all}
\end{figure*}

Next, we further evaluate the stage results of reconstructed facades on Bridge-1 example. The visualization of the aerial flight path, the facade geometry, the facade proxy, and facade texture are shown in Fig.~\ref{fig:stage-results-all}. Specifically, we visualize the errors on geometry and texture compared to GT as color maps.  
The results still indicates that the reconstructed geometry of DS and our method is more accurate that that of OP. The error maps on textures exhibit large differences among OP, DS and our method. The results of DS and ours achieves higher frontality compared to OP. And our method and DS plan views closer to the buildings, the texture maps exhibits much higher sharpness, outperforming the OP method. Meanwhile, our method achieves texture maps with the least errors.

In summary, for structured proxy reconstruction, the key differences among the methods lie in texture quality and acquisition cost. Therefore, we focus our subsequent comparisons on these two aspects.

\subsection{Comparison on Virtual Scenes}

\begin{figure*}
	\centering
	\includegraphics[width=\linewidth]{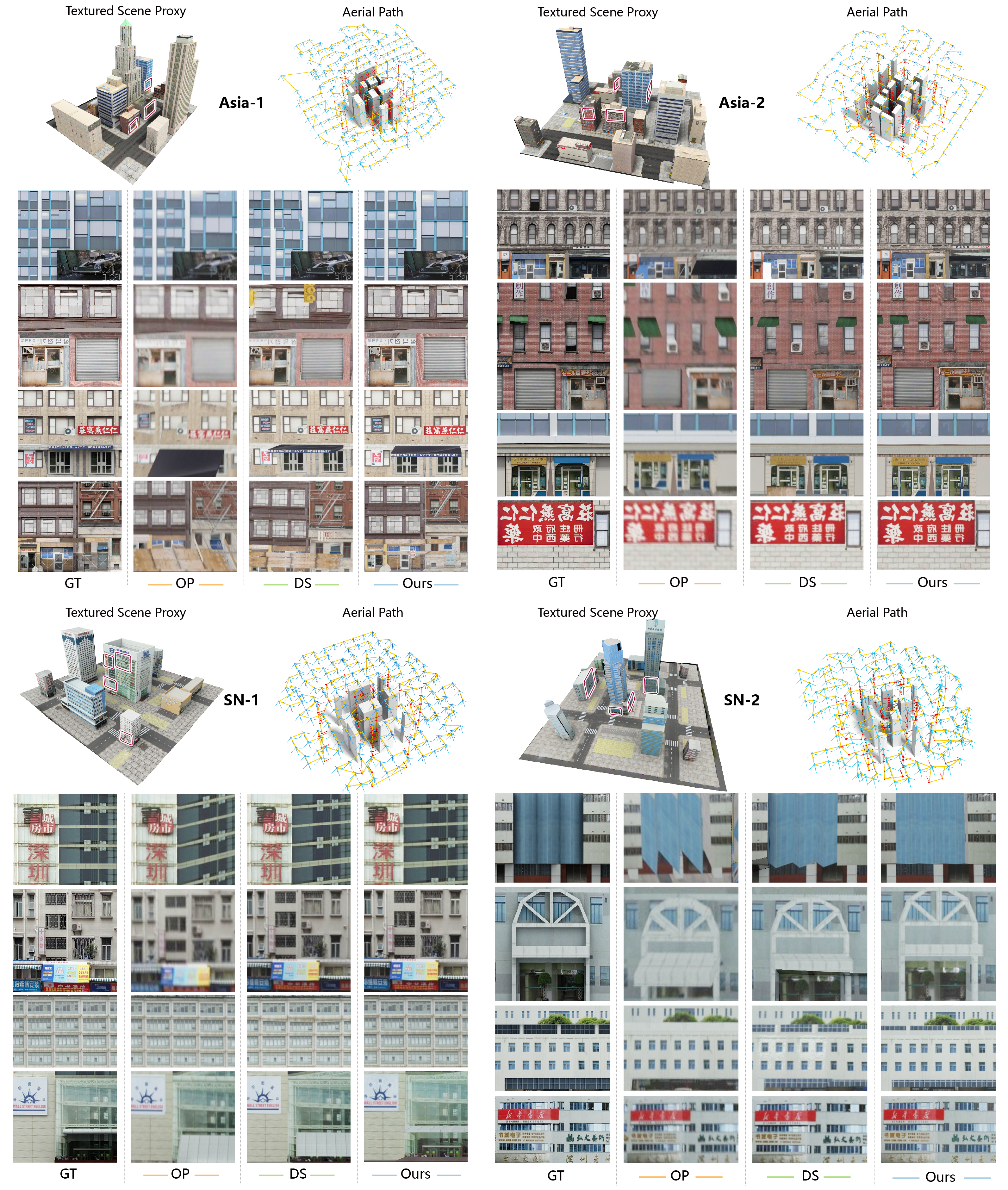}
	\caption{
		Visual comparison against reconstructed texture maps with GT, OP, DS, and ours on two synthetic scenes. Detailed comparisons of textured proxy models are shown in the zoomed-in insets.
	}
	\label{fig:comparison2}
\end{figure*}

\paragraph{Qualitative evaluation.}
Fig.~\ref{fig:comparison2} shows the qualitative comparison results on four synthetic scenes with the ground truth texture maps (GT).
First, without enforcing constraints on perspective quality during view selection, both OP and DS fail to consistently capture fronto-parallel images, resulting in noticeable artifacts such as stretching and perspective inconsistencies in the generated textures. The stretching problem, caused by using oblique views, is evident in the inset of Asia-1 scene (third and fourth rows, second and third columns), Asia-2 scene (second row, sixth column), SN-1 scene (fifth and eighth row, second and third columns) and SN-2 scene (fifth and sixth rows, sixth column) in Fig.\ref{fig:comparison2}. This issue commonly arises when texturing facades using images collected with OP. 
The perspective inconsistency can be observed in the insets of the SN-2 example in Fig.~\ref{fig:comparison2} (seventh row, seventh column), which is common when texturing ground or roofs with photos collected by DS.
In contrast, our planning method considers both front-parallel and perspective consistency, providing scene surfaces with photos of high perspective quality. 
Both our method and DS plan views at low altitudes, close to the building facades. As a result, the texture maps produced by DS and our method exhibit significantly higher resolution than those from OP, with our method consistently yielding slightly higher resolution than DS, this pattern is observable across all tested scenes. 
Furthermore, our method and OP produce texture maps with much less structural distortion. This advantage stems from capturing images from a greater distance, allowing large facade areas to be covered by very few photos. Such distortions can be found in the insets of the Asia-1 example in Fig.~\ref{fig:comparison2} (first and second rows, third column), and the SN-2 example in Fig.~\ref{fig:comparison2} (seventh row, seventh column).

\begin{table*}
	\caption{Quantitative comparison and statistics on the planning and texturing of virtual scenes. Numbers in the bracket denote the quality value of the four zoomed-in views of each example from top to bottom.}
	\label{tab:quality-comparison1}
	\begin{center}
		\begin{tabular}{c c c c c c c c c}
			\midrule[1pt]
			\multirow{2}{*}{Scene} & \multirow{2}{*}{Method} & \multicolumn{3}{c}{Capturing} & \multicolumn{2}{c}{Textured Proxy}   \\
			\cmidrule(lr){3-5} \cmidrule(lr){6-7}
			&   & $\#$ Images & $\#$ Hover  & Trajectory & $ \small{SSIM}\uparrow$ & $\small{LPIPS} \downarrow$   \\
			\cmidrule(lr){1-1} \cmidrule(lr){2-2} \cmidrule(lr){3-5} \cmidrule(lr){6-7}
			\multirow{3}{*}{Asia-1}        & OP   & 1615   & 323  & \numprint{8653} 	  & (0.443, 0.527, 0.417, 0.381)  & (0.666, 0.779, 0.730, 0.803)    \\
			\multicolumn{1}{c}{}        & DS   & 2420 & 2420  & \numprint{25091}   & (0.479, 0.487, 0.316, \textbf{0.455})&  (0.359, 0.527, 0.560, 0.590)    \\
			\multicolumn{1}{c}{}        & Ours & 1653   & 495   &  \numprint{12224} &  (\textbf{0.486}, \textbf{0.575}, \textbf{0.493}, 0.387)&  (\textbf{0.309}, \textbf{0.213}, \textbf{0.132}, \textbf{0.436})  \\
			\cmidrule(lr){1-1} \cmidrule(lr){2-2} \cmidrule(lr){3-5} \cmidrule(lr){6-7}				
			\multirow{3}{*}{Asia-2}        & OP   & 1200  & 240   & \numprint{9201}  & (\textbf{0.336}, \textbf{0.409}, \textbf{0.584}, 0.561)     & (0.791, 0.804, 0.501, 0.722)     \\
			\multicolumn{1}{c}{}        & DS   & 2011 & 2011   & \numprint{26501} &(0.289, 0.342, 0.563, 0.618)      &  (0.582, 0.528, 0.374, 0.555)  )   \\
			\multicolumn{1}{c}{}        & Ours & 1370   & 440   & \numprint{15466} &(0.272, 0.382, 0.567, \textbf{0.636})      &  (\textbf{0.351}, \textbf{0.229}, \textbf{0.271}, \textbf{0.336})     \\
			\cmidrule(lr){1-1} \cmidrule(lr){2-2} \cmidrule(lr){3-5} \cmidrule(lr){6-7}			
			\multirow{3}{*}{SN-1}        & OP   & 1125  & 225   & \numprint{4783} &  (0.391, 0.369,\textbf{0.729}, 0.550)  & (0.426, 0.680, 0.151, 0.399)     \\
			\multicolumn{1}{c}{}        & DS   & 958 &  958  & \numprint{10625}   &(0.455, 0.415, 0.603, 0.587) &  (0.337, 0.386, \textbf{0.114}, 0.203)    \\
			\multicolumn{1}{c}{}        & Ours & 1113   & 304   & \numprint{5781}  &(\textbf{0.503}, \textbf{0.448}, 0.699, \textbf{0.654})   &  (\textbf{0.295},\textbf{ 0.305}, 0.158,\textbf{ 0.167})   \\
			\cmidrule(lr){1-1} \cmidrule(lr){2-2} \cmidrule(lr){3-5} \cmidrule(lr){6-7}			
			\multirow{3}{*}{SN-2}        & OP   & 1125  &  225  & \numprint{4783}  &(0.603, 0.727, 0.603, 0.641)    &  (0.528, 0.474, 0.497, 0.436)     \\
			\multicolumn{1}{c}{}        & DS   & 827 &  827  & \numprint{9874}  &(0.695, 0.764, \textbf{0.696}, 0.647)     &   (0.394, 0.306, 0.381, 0.297)    \\
			\multicolumn{1}{c}{}        & Ours & 1136   & 323   & \numprint{5883} & (\textbf{0.756}, \textbf{0.791}, 0.627, \textbf{0.846})     &  (\textbf{0.268}, \textbf{0.278}, \textbf{0.327}, \textbf{0.122})    \\
			\midrule[1pt]
		\end{tabular}
	\end{center}
\end{table*}

\begin{figure*}
	\centering
	\includegraphics[width=\linewidth]{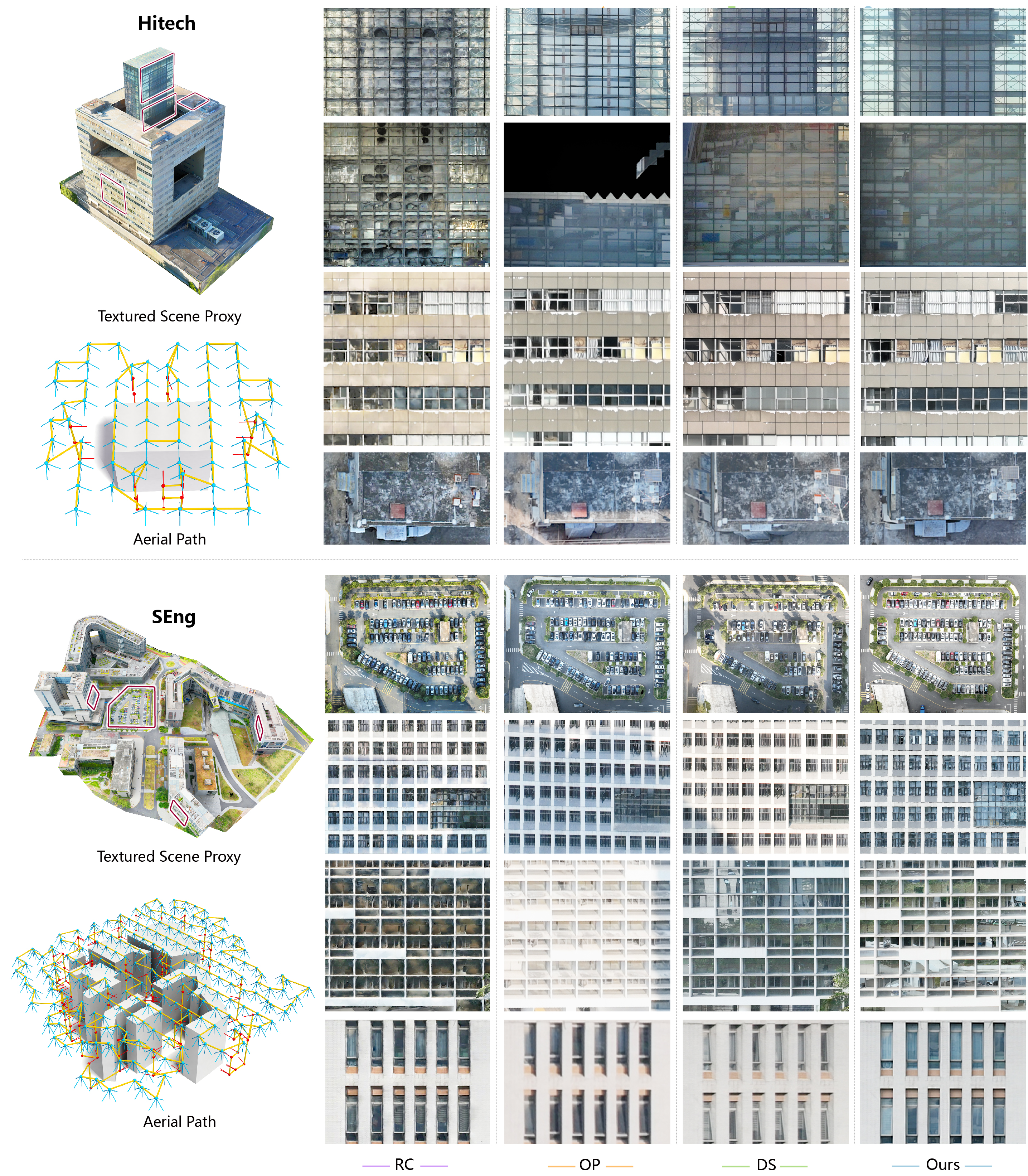}
	\caption{
		Visual comparison against reconstructed texture maps with RC, OP, DS, and ours on two real-world scenes. The detailed comparisons are shown in the zoomed-in insets.}
	\label{fig:comparison-real}
\end{figure*}

\begin{figure*}
	\centering
	\hspace{2mm}\includegraphics[width=.98\linewidth]{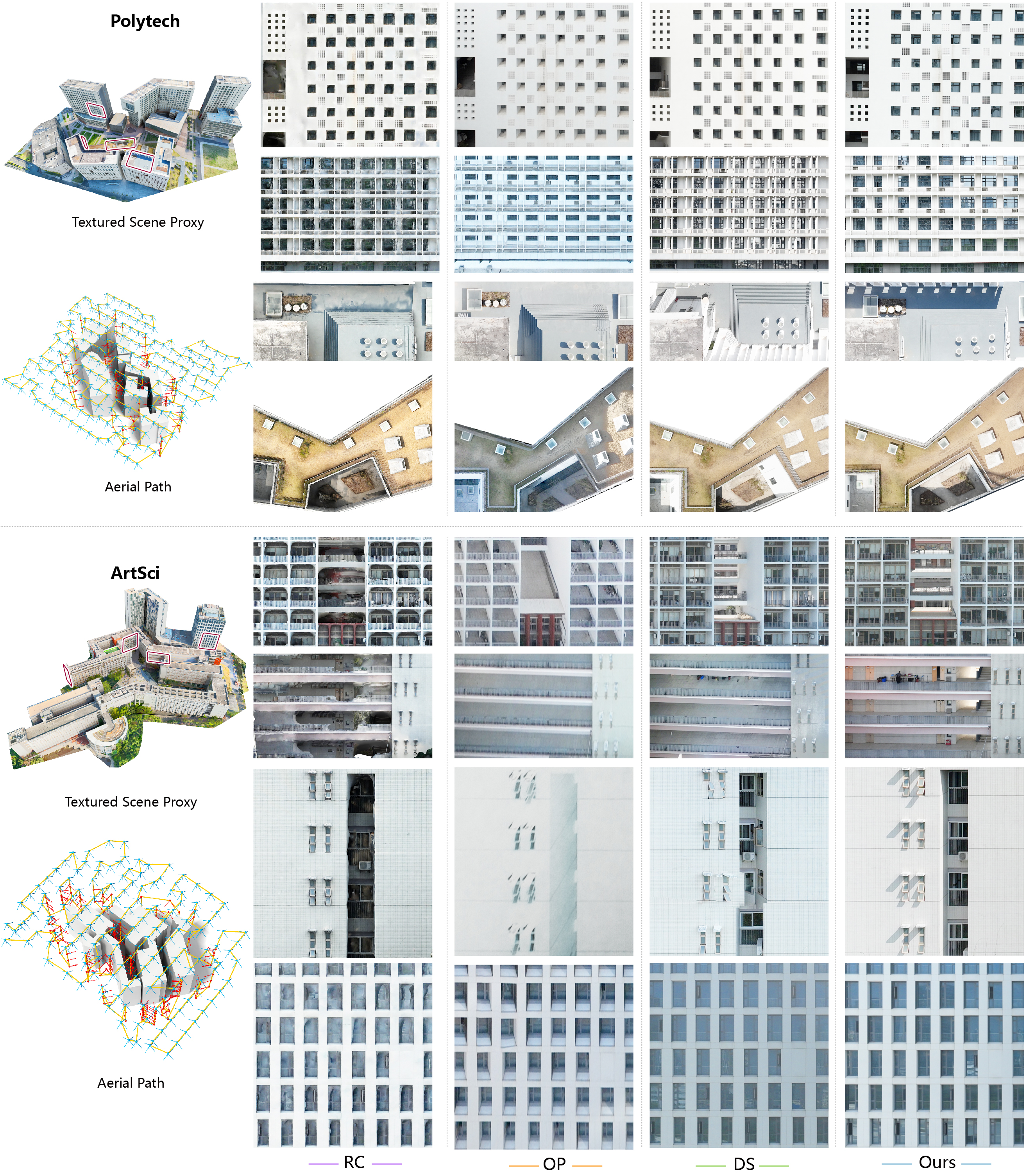}
	\caption{
		Visual comparison against reconstructed texture maps with RC, OP, DS, and ours on two real-world scenes. The detailed comparisons are shown in the zoomed-in insets.}
	\label{fig:comparison-real2}
\end{figure*}

\begin{table*}
	\caption{Quantitative comparison and statistics on the planning, capturing and textured proxies of real scenes. Numbers in the bracket denote the quality value of the four zoomed-in views of each example from top to bottom.}
	\label{tab:quality-comparison2}
	\begin{center}
		\resizebox{1.0\linewidth}{!}{%
			\begin{tabular}{c c c c c c c c c c}
				\midrule[1pt]
				\multirow{2}{*}{Scene} & \multirow{2}{*}{Method}  & \multirow{2}{*}{Time Slot}   & \multirow{2}{*}{$\#$ Images}  & \multirow{2}{*}{$\#$ Hover}   & Trajectory & Energy & Time & \multirow{2}{*}{$ \small{SSIM}\uparrow$} & \multirow{2}{*}{$\small{LPIPS} \downarrow$}  \\
				& & & & & (m) &  (\%) &  (min) & & \\
				\cmidrule(lr){1-1} \cmidrule(lr){2-2} \cmidrule(lr){3-8} \cmidrule(lr){9-10}
				\multirow{3}{*}{Hitech}       & OP   & 14:30-15:00   & 360 & 72 & \numprint{2085} & 44 & 13   &   (0.308, 0.198, 0.622, 0.222 )     &  (0.440, 0.575, \textbf{0.337}, 0.563 )      \\
				\multicolumn{1}{c}{}        & DS   & 15:00-16:30   & 567 & 567 & \numprint{7629} & 214 & 76   &  (0.370, 0.273, 0.575, \textbf{0.246} )     &  (0.403, \textbf{0.307}, 0.371, \textbf{0.514} )     \\
				\multicolumn{1}{c}{}        & Ours  & 16:30-17:00   & 347  & 87 & \numprint{2806} & 51 & 15   &  (\textbf{0.395}, \textbf{0.299}, \textbf{0.627}, 0.221 ) &(\textbf{0.337}, 0.324, 0.378, 0.524)    \\
				\cmidrule(lr){1-1} \cmidrule(lr){2-2} \cmidrule(lr){3-8}  \cmidrule(lr){9-10}
				\multirow{3}{*}{ArtSci}       & OP    & 9:30-10:30   & 485 & 97 & \numprint{4908} & 103 & 31  &   (0.200, 0.306, 0.147, 0.529 )     &  (0.590, 0.734, 0.575, 0.334 )        \\
				\multicolumn{1}{c}{}        & DS    & 11:00-16:30   & 1722 & 1722 & \numprint{20498} & 825 & 241  &  (0.284, 0.331, 0.338, 0.585 )     &  (0.375, 0.732, 0.271, 0.268 )         \\
				\multicolumn{1}{c}{}        & Ours  & 10:30-12:30  & 1121 & 302 & \numprint{9737} & 206 & 63  & (\textbf{0.289}, \textbf{0.398}, \textbf{0.586}, \textbf{0.613} )     &  (\textbf{0.323}, \textbf{0.666}, \textbf{0.201}, \textbf{0.186} )   \\	
				\cmidrule(lr){1-1} \cmidrule(lr){2-2} \cmidrule(lr){3-8} \cmidrule(lr){9-10}	 	
				\multirow{3}{*}{Polytech}   & OP   & 14:00-15:00   & 1200  & 240 & \numprint{6261} & 140 & 45  &    (0.512, 0.264, 0.389, 0.072)              &  (\textbf{0.600}, 0.530, 0.582, 0.582)          \\
				\multicolumn{1}{c}{}        & DS   & 10:00-14:00   & 1337 & 1337 & \numprint{18521} & 530 & 162  &    (0.398, 0.319, 0.403, 0.467)               &  (0.606, 0.542, 0.583, 0.535)    \\
				\multicolumn{1}{c}{}        & Ours  & 16:00-17:30  & 1252 & 334 & \numprint{10499} & 212 & 65   & (\textbf{0.487}, \textbf{0.367}, \textbf{0.559}, \textbf{0.590})     &  (0.607, \textbf{0.520}, \textbf{0.445}, \textbf{0.482})    \\
				\cmidrule(lr){1-1} \cmidrule(lr){2-2} \cmidrule(lr){3-8} \cmidrule(lr){9-10}
				\multirow{3}{*}{SEng}       & OP    & 16:30-17:30  & 1100 & 220 & \numprint{5960} & 123 & 38  &   (0.512, 0.676,0.511, 0.255 )     &  (0.654,0.585,0.631, 0.590 )    \\
				\multicolumn{1}{c}{}        & DS    & 13:00-17:00  & 1850 & 1850 & \numprint{26248} & 783 & 217  &  ( 0.563, 0.584, 0.554, \textbf{0.378})     &  (0.650, 0.637, 0.630, 0.578 )     \\
				\multicolumn{1}{c}{}        & Ours  & 9:30-11:30   & 1559 & 417  & \numprint{11710} & 230 & 74   & (\textbf{0.604}, \textbf{0.687}, \textbf{0.611}, 0.274 )     &  (\textbf{0.638}, \textbf{0.520}, \textbf{0.600}, \textbf{0.550} )     \\
				\midrule[1pt]
			\end{tabular}
			}
	\end{center}
\end{table*}

\paragraph{Quantitative evaluation.}
To evaluate the effectiveness of our co-captured photo collections in texture reconstruction, we compare rendered images of ground truth (GT) facades with those of reconstructed facades, all viewed from a fronto-parallel perspective.

The comparison results are reported in Table~\ref{tab:quality-comparison1}. SSIM tends to be less sensitive to blurring artifacts and may assign relatively high scores to images affected by regional blurring~\cite{zhang2018unreasonable}. This phenomenon can be observed in some views per each example. Meanwhile, LPIPS is designed to match human perception and yields better scores for images with a higher level of coherence.  Across most tested cases, our reconstructed textures consistently achieve higher scores in both SSIM and LPIPS, indicating superior visual quality and perceptual fidelity in the rendered results.

\paragraph{View number comparison.}
The number of planned views has a direct impact on the total flight path length and, consequently, the energy consumption~\cite{DroneScan20}. We compare the number of views generated by OP, DS, and our method across four virtual scenes. As reported in Table~\ref{tab:quality-comparison1}, our method and OP produce a similar number of views. In contrast, DS tends to generate more views for scenes characterized by tall buildings or higher building density, such as Asia-1 and Asia-2. Conversely, for scenes with shorter or fewer buildings (such as Bridge-1, C-1, SN-1, and SN-2), DS plans fewer views. For all the scenes, OP delivers the shortest flight and least number of hovering positions. Meanwhile, DS plans flights with 156\%-380\% more hovering positions and 68\%-105\% longer trajectories compared to our method. According to previous experiments, these indicates that DS requires the longest capturing time and energy cost.

\subsection{Comparison on Real Scenes}

All photos of real-world scenes are captured using a DJI MAVIC 3E drone. 
{A wide-angle lens was used, following common practice in aerial photography, as it is less sensitive to platform vibrations and provides a larger field of view that benefits feature mapping.}
In practice, environmental factors such as weather can cause significant visual variations, particularly in color tone, even when photos are taken at the same time on different days. To ensure a fair comparison of reconstructed textures, we make every effort to capture the same scene under optimal daylight conditions on the same day. Furthermore, to eliminate bias related to capture order, we assign a randomized execution sequence to each method (see Table~\ref{tab:quality-comparison2}).

\paragraph{Qualitative comparison.}
Fig.~\ref{fig:comparison-real} and Fig.~\ref{fig:comparison-real2} show the qualitative results on four real scenes.
We provided the dense model of facades reconstructed with RC as the reference for ground truth color and structure.
The texture maps produced by DS and our method exhibit much higher resolution than OP, with our method achieving slightly higher resolution than DS across all examples. 
In terms of perspective quality, both OP and DS fail to capture images with consistently front-parallel viewing directions, resulting in noticeable artifacts. The stretching problem can be observed in the inset of SEng (sixth and eighth rows, second and third columns) in Fig.~\ref{fig:comparison-real}, Polytech (third and fourth rows, third column) and ArtSci (fifth to seventh rows, second and third columns) in Fig.~\ref{fig:comparison-real2}. Perspective inconsistencies are also evident in the SEng example (fifth row, third column) in Fig.~\ref{fig:comparison-real}, in Polytech (third row, second column) and ArtSci (seventh row, third column and eighth row, second column) in Fig.~\ref{fig:comparison-real2}. 
In contrast, our planning method accounts for both front-parallel and perspective consistency, yielding superior texture quality, particularly for ground and roof surfaces where DS underperforms. This advantage is demonstrated in the SEng scene (fifth row) in Fig.~\ref{fig:comparison-real} and the Polytech example (third and fourth rows) in Fig.~\ref{fig:comparison-real2}. 
Additionally, our method produces texture maps with significantly reduced structural distortion. For instance, such distortions are visible in the Hitech example (third row, third column) in Fig.~\ref{fig:comparison-real}, and ArtSci example (seventh row, third column) in Fig.~\ref{fig:comparison-real2}, whereas our method effectively mitigates them.

\paragraph{Quantitative comparison.}
We adopt the evaluation scheme in~\cite{TwinTex23} and~\cite{waechter2014let} to compare rendered images of the reconstructed scenes against the corresponding real images following a specific view from the input cameras.
We select the ground truth photos for evaluation from the input excluding the photos for texturing.

Quantitative comparison results are presented in Table~\ref{tab:quality-comparison2}. 
SSIM is inherently less sensitive to blurring artifacts and may yield higher scores for images with such issues~\cite{zhang2018unreasonable}. This phenomenon is observed in several views across the examples. Meanwhile, in most cases, the rendered views of our textured models achieve superior SSIM, LPIPS, and $Q_s$ scores.

\paragraph{Energy cost and view number comparison.}
For battery-operated drones, energy consumption is a critical constraint that directly impacts operational duration. To reduce energy consumption, \citet{roberts2017submodular} focused solely on minimizing flight path length, while \citet{DroneScan20} also considered the number of captured views. 
The statistics of OP, DS and our method on photo capturing for all the scenes are provided in Table~\ref{tab:quality-comparison2}.
Unlike virtual scenes, even if the real scenes have various building height and densities, we found that DS requires a larger number of views to achieve feasible reconstruction for all the real scenes. 
While OP incurs the lowest cost, our method significantly reduces resource usage compared to DS. Specifically, our approach requires only 61\%-94\% of the images, achieves a 15\%-25\% reduction in the number of hovering positions, shortens the trajectory length by 37\%-48\%, and lowers energy consumption by 24\%-40\%.

In summary, the results demonstrate several clear advantages of our approach: i) higher level of front-parallel and perspective consistency, ii) higher photometric quality (\ie resolution), iii) minimal distortion, blurring, or seaming artifacts from multi-view blending, iv) efficient acquisition of high-quality photos for geometry and texture with low energy consumption.

\begin{figure*}
	\centering
	\includegraphics[width=\linewidth]{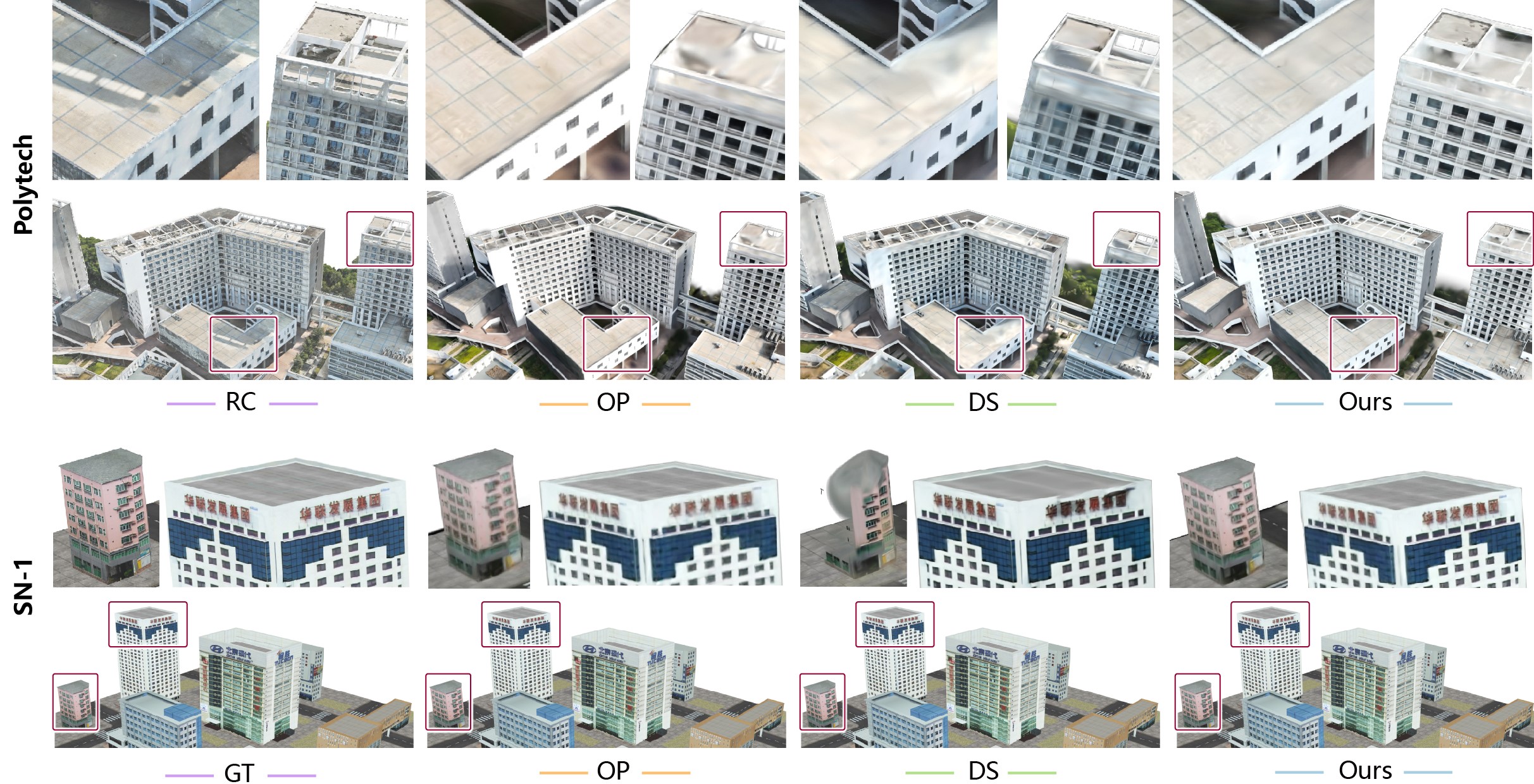}
	\caption{
		Visual comparison against reconstructed 2DGS representation with photos captured according to OP, DS, and ours on a real-world scene (top) and a synthesis scene (bottom). The mesh models of RC and GT are provided to serve as reference for ground truth scene structure.}
	\label{fig:comparison-implicit}
\end{figure*}

\subsection{Results on Learning-based Reconstruction}

Finally, we compare the reconstructed 2DGS~\cite{huang20242d} using images captured by our method, OP, and DS on both a synthetic and a real-world scene. The evaluation reveals several key observations: i) When using COLMAP~\footnote{https://colmap.github.io/} to calibrate the captured images~\cite{huang20242d}, the calibration often fails to estimate accurate camera poses for images planned with DS, particularly in synthetic scenes, resulting in severe artifacts such as ghosting (\eg third row, third column of Fig.~\ref{fig:comparison-implicit}). In contrast, COLMAP performs reliably with OP and our method. This discrepancy is primarily due to DS's design, which emphasizes close-up captures of facades, producing repetitive and similar patterns that have adverse effect on calibration; ii) OP produces fewer floaters than our method, and both OP and our approach outperform DS in this regard. The reduction in floaters is attributed to minimized sky observations. For fair comparisons, floaters outside the building volume bounding box were removed wherever possible.

Since the reconstructed scenes are not in a shared coordinate space, we carefully selected similar and previously unseen camera angles for visualization in Fig.~\ref{fig:comparison-implicit}. For the synthetic scene, ground truth (GT) is provided; for the real scene, a high-quality dense reconstruction by RC is used as reference. The results show that both OP and our method generate clearer details on roofs and large open grounds (first row in Fig.~\ref{fig:comparison-implicit}), whereas DS better preserves facade details, especially at lower altitudes. Both DS and our method outperform OP in capturing structured facade geometry, and DS achieves the most detailed facades in close-range views.
Overall, our method offers the most consistent reconstruction quality across the entire scene.

\section{Conclusion and Future Work}
The texture quality is often sacrificed in automated systems to get dense geometry. In this work, we introduce the problem of geometry and texture co-capture. Given limited input information, we propose a novel aerial path planning algorithm that enables the reconstruction of both geometric proxy models and realistic texture maps for urban scenes.
Two quality metrics are introduced given 2D map and 2D views: one is to measure the quality of a building facade in a complex scene given a set of 2D views, and the other one is to measure the quality of a view related to a building facade.
Based on these metrics, a 3D view planning algorithm is implemented through multi-objective optimization approaches, that simultaneously enhances texture fidelity and reduces the costs associated with aerial image acquisition.
We evaluated our method using several synthetic and real-world urban scenes and demonstrated its advantages by comparing it to previous methods. 
To our knowledge, this is the first work that explicitly addresses geometry–texture co-capture. Moreover, our simple yet practical solution achieves a strong balance between reconstruction quality and other key goals of automation, including consistent lighting, operation in confined spaces, flight safety, acquisition speed, and limited input information.

\begin{figure}
	\centering
	\includegraphics[width=.85\linewidth]{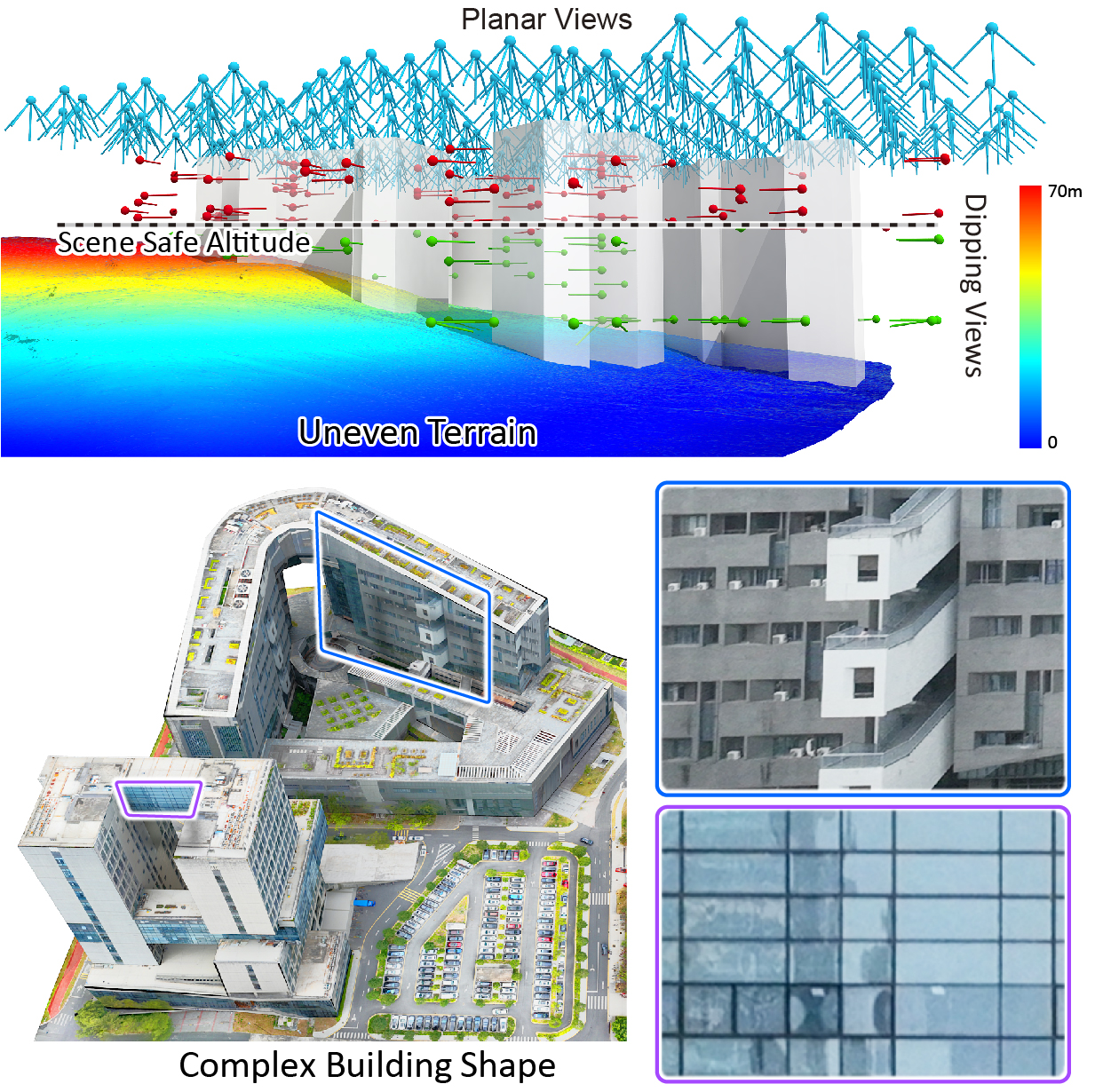}
	\caption{{Top: 3D views of a terrain-located scene. Without accurate terrain altitude information, we can only create the red dipping views above the scene safe altitude (black dashed line). Given sufficient information, additional green dipping views can be placed below the safe altitude. Terrain altitude is color-coded from blue (low) to red (high). Bottom: Inner facades of complex buildings exhibiting less clear, less front-parallel textures, caused by large discrepancies between their 3D shapes and 2D silhouettes.}}
	\label{fig:limitations}
\end{figure}

\paragraph{Limitations and future work.}
First, since optimization accounts for most of the runtime, we plan to accelerate this step through parallel processing.
Second, we assume that the exact altitude of the target area is unknown, and only a scene safe altitude is given (black dashed line in Fig.~\ref{fig:limitations} (top)). In our current implementation, all dipping sequences share the same minimum altitude. For a scene located on uneven terrain, only the dipping views above the safe altitude are generated, see the red views in Fig.~\ref{fig:limitations} (top). If accurate altitude information becomes available, the vertical dipping range can be dynamically adjusted, allowing additional dipping views below the safe altitude (see the green views in Fig.~\ref{fig:limitations} (top)). We regard this as a future engineering extension. Similarly, because our input is limited to a 2D map with approximate building contours, highly concave upper/inner facades may receive lower-resolution textures and more oblique viewing angles (see Fig.~\ref{fig:limitations}, bottom). A potential solution is to analyze more detailed 2D contours at multiple altitudes and plan dipping points accordingly. Furthermore, if accurate building heights are given, photo capture could be reduced by setting $H$ to the building height during vertical view generation, while keeping other steps unchanged. 
Third, TwinTex~\cite{TwinTex23} represents the current state of the art in architectural proxy texturing. When proxy accuracy falls within its tolerance range, the results are plausible; otherwise, facade seams may appear on the facade boundary. These seams mainly stem from (i) camera calibration errors, (ii) errors introduced in model abstraction, and (iii) limitations of the texturing step. High-quality photos, featuring large views, clear content, and sufficient overlap, can mitigate such artifacts. As demonstrated in our video, texture maps generated with our photos contain substantially fewer artifacts. We regard improvements to the sub-steps of the reconstruction pipeline as future work.
Fourth, while view-dependent reflectance, relightability, and material estimation can be achieved via physically based rendering (PBR), such estimation lies outside our current scope since our problem centers on geometry–texture co-capture. Users requiring PBR properties may derive them from our captured photos or reconstructed textures, either by manually assigning materials in off-the-shelf graphics engines or by applying learning-based methods (e.g., RGB$\leftrightarrow$X~\cite{zeng2024rgb}) to infer PBR materials. 
Finally, reconstructions from photos captured by OP, DS, and our method all show foliage around similar facade regions, as these areas are consistently occluded in the input photos. For users preferring foliage-free reconstructions, similar to those in commercial maps, post-processing can be applied to the texture maps, for instance, foliage segmentation (e.g., PSPNet~\cite{zhao2017pyramid}) followed by image inpainting (e.g., the inpainting model in TwinTex~\cite{TwinTex23}).

\section*{Acknowledgments}
We thank the reviewers for their constructive comments. This work was supported in parts by National Key R\&D Program of China (2024YFB3908500, 2024YFB3908502), NSFC (U21B2023, 62302313, 62572059), Guangdong Basic and Applied Basic Research Foundation (2023B1515120026), Shenzhen S\&T Program (KJZD20240903100022028, KQTD20210811090044003, RCJC20200714114435012), and Scientific Development Funds from Shenzhen University.

\bibliographystyle{ACM-Reference-Format}
\bibliography{DroneTex}
\end{document}